\documentclass{iopart}
\usepackage{graphicx}
\usepackage{amssymb}
\begin{document}

\def\dirac{i\partial\!\!\!/}
\def\Dirac{iD\!\!\!\!/}

\def\P{P\!\!\!\!/}
\def\DDirac{i\partial\!\!\!/+eA\!\!\!/}
\def\EDirac{D\!\!\!\!/\,}
\def\e{{\rm e}}
\def\egamma{\buildrel-\over\gamma}
\def\eE{\buildrel-\!-\over E}
\def\ex{\buildrel-\over x}
\def\eA{\buildrel-\over A}
\def\oq{\buildrel-\over q}
\def\pa{\partial}
\def\rmd{{\rm d}}
\def\rmi{{\rm i}}
\def\vu{\upsilon}
\def\ve{\varepsilon}
\def\sumint{\sum\!\!\!\!\!\!\!\!\int}
\def\lbar{^-\!\!\!\!\lambda}
\def\R{\Re{\rm e}}
\def\I{\Im{\rm m}}

\newcommand{\beq}{\begin{eqnarray}}
\newcommand{\eeq}{\end{eqnarray}}
\newcommand{\nn}{\nonumber}
\newcommand{\no}{\nonumber\\}

\article[Canonical quantization with background fields]{Canonical quantization with background fields}
{Pairs Emission in a Uniform Background Field:\\ an Algebraic Approach}

\date{\today}

\author{by Roberto Soldati}\address{Dipartimento di Fisica - Universit\`a
di Bologna, Istituto Nazionale di Fisica Nucleare - Sezione di
Bologna}
\begin{abstract}
A fully algebraic general approach is developed to treat the pairs emission and absorption
in the presence of some uniform external background field. In particular, it is shown that
the pairs production and annihilation operators, together with the pairs number operator,
do actually fulfill the SU(2) functional Lie algebra. As an example of application,
the celebrated Schwinger formula is consistently
and nicely recovered, within this novel approach, for a Dirac spinor field 
in the presence of a constant and homogeneous electric field in four spacetime dimensions.
\end{abstract}
\section{Introduction}
Soon after the discovery of the relativistic Dirac equation \cite{Dirac}
it was quickly realized \cite{sauter, euler} that electron positron pairs might have been
produced out of the vacuum by e.g. a constant homogeneous very strong electric field.
Later on Julian Schwinger succeeded in retrieving \cite{schwinger}
the electron positron pairs creation rate
per unit time and volume, by exploiting the proper time method and the analytic continuation from the
four dimensional Minkowski spacetime to the euclidean space.
Since then, on the one hand, a huge amount of work, papers and monographies on this subject
has been put forward \cite{IZ,review,dunne}. On the other hand, the experimental probe 
of the famous Schwinger formula for the electron positron pairs creation rate
is still lacking nowadays, because of the very large value of the 
corresponding electric field strength critical scale \cite{dunnegies}.
Nonetheless, it has been quite recently suggested \cite{graphene} that
in the pseudorelativistic planar QED effective model for graphene,
the detectable emission of massless Dirac quasiparticles antiquasiparticles pairs
might occur.

Surprisingly enough, it was only after nearly twenty years, since the
seminal Schwinger article, that the exact solutions of the Dirac equation in the presence
of a uniform electric field
was obtained by Nikishov \cite{nikishov}. In this remarkable paper
\footnote{I am truly indebted to Gerald V. Dunne for bringing this absolutely relevant paper to my attention.}
as well as within some subsequent ones \cite{nikishov2}, arguments have been put forward to
reobtain the Schwinger formula for the electron positron pairs creation rate
from the knowledge of the nonperturbative exact solutions of the Dirac equation
in a constant and homogeneous electric field. 
However, although physically and heuristically well posed, those arguments do not appear, at least in my opinion,
to be completely developed and fully convincing from the modern field theoretical point of view.
It is the main target of the present paper to fill this gap.

It is worthwhile to remark that those exact
solutions of the Dirac equation are nonstationary. This is because of the gauge choice,
 which renders the problem exactly
solvable, but is such that the ensuing Dirac hamiltonian becomes time dependent.
Moreover, those exact solutions are truly nonperturbative in the sense that
there is no asymptotic regime in which they reduce to the free field plane wave solutions,
just because a uniform electric field is never negligible.
As a consequence, it turns out that the asymptotic states and fields can never
be identified with the conventional ones \footnote{In the literature,
unfortunately,
there are several authors that roughly and wrongly disregard this crucial feature
and come unavoidably to incorrect conclusions.},
which are solutions of a the free field theory
without external fields.

It is the aim of the present paper to reformulate the pairs emission and absorption
issue in a purely algebraic context based upon the canonical quantization
in the presence of background, external, uniform fields.
This algebraic approach is quite general for it does not rely on
the specific form of the Dirac equation and of the uniform field.
For example, it might be used to describe matter production in a 
constant gravitational field, or the Unruh effect \cite{unruh}
in a Rindler spacetime. 

The main feature of this approach lies in the fact
that the pairs creation, the pairs destruction and the pairs number operator
do realize a representation of the functional SU(2) Lie algebra.
This idea is not new, since it dates back to the seminal paper by
Marinov and Popov \cite{marinov}, in which a group theoretical
approach to the problem of pair creation in a time dependent
electric field has been developed. However, it is my aim to fully develop 
this approach, in the modern language of quantum field theory and in a rigorous
way, taking the massive Dirac field in a constant electric field as a paradigmatic example.
This allows in turn to understand the Bogolyubov transformations \cite{NNB}
as similarity transformations acting upon the Fock space of the states,
while the vacuum state will correspond
to a coherent state {\em \`a la} Dirac \cite{Dirac} with an infinite
sea of pairs. In so doing, the Schwinger formula will nicely appear to be
the vacuum expectation value of an operator which
represents a complex functional rotation around a fixed axis.

For the sake of clarity and to provide a paradigmatic example,
in the first two sections
I will provide a short overview for the solutions of the Dirac equation
in the presence of a constant homogeneous electric field on the Minkowski
spacetime, as well as the canonical quantization of the Dirac field
and the ensuing derivation of the Schwinger formula.
In so doing, I will establish my notations and conventions.
Then in section \ref{algebraic} the algebraic approach will be developed
and all the formulas will be checked in terms of the four dimensional
spinor QED with a uniform electric field.
Finally, a short discussion of the concluding remarks will be attempted.
\section{Exact Solutions of the Dirac Equation}
In this Section I briefly review the structure of the nonstationary (time dependent)
exact (nonperturbative) solutions of the Dirac equation, in the presence of a constant and homogeneous electric
field on the four dimensional Minkowski spacetime.
I will use the metric $g_{\,\mu\nu}={\rm diag}\,[\,+, -, -, -\,]\,,$
natural units $\hbar=c=1$ and the ordinary or standard 
representation for the Dirac matrices \cite{landaulifsits}.
The Dirac equation can be written either in the covariant form
\beq
(\DDirac -M)\Psi(x)
=(\gamma^\mu P_\mu-M)\Psi(x)=0
\label{eeeqdirac}
\eeq
where $q=-\,e\ (e>0)$ is the negative electron charge and 
$p^{\,\mu}=i\partial^{\,\mu}=(i\partial_{\,t}\,,-\,i\nabla)$
while $P_{\,\mu}=p_{\,\mu}- qA_{\,\mu}(x)=i\partial_{\,\mu} + eA_{\,\mu}(x)$
is the usual abelian covariant derivative, or {\it \`a la} 
Schr\"odinger, viz.,
\beq
i\,{\partial\Psi\over \partial t}= H\,\Psi\,,\qquad\quad
H =-eA_0 + \alpha^k P^{\,k} + \beta\,M
\eeq
in which we have set as it is customary
\[
\beta\equiv\gamma^0\qquad\quad
\gamma^0\gamma^k\equiv\alpha^k\qquad\quad 
{\bf P} = -\,i\,\nabla+e{\bf A}
\]
If we assume the electrostatic field towards the positive 
$OX$ axis, that means $F^{10}=F_{01}=E_x=E>0\,,$ after setting 
$x^{\,\mu}\equiv(t,{\bf r})=(t,x,y,z)\,,$
we get the {\sl time dependent} hamiltonian in the Nikishov \cite{nikishov} temporal gauge
\footnote{Actually, the Nikishov temporal gauge is a particular case of the general
Fock-Schwinger gauge choice $A_{\,\mu}=\frac12\,F_{\,\rho\mu}\,x^{\,\rho}\,.$}
$A^\mu=(0,-Et,0,0)$
\beq
H_t = -(i\partial_x+eEt)\,\alpha^1-i\partial_y\,\alpha^2
-i\partial_z\,\alpha^3 + M\beta
\eeq
which turns out to be a symmetric operator but does not allow for stationary states.
It follows that  the first order Dirac operators read
\beq
\fl\qquad
\DDirac \pm M =
\left\lgroup\begin{array}{cccc}
i\partial_t \pm M& 0& i\partial_z& iD_x+\partial_y\\
0& i\partial_t \pm M& iD_x-\partial_y& -i\partial_z\\
-i\partial_z& -iD_x-\partial_y& -i\partial_t\pm M& 0\\
-iD_x+\partial_y& i\partial_z& 0& -i\partial_t\pm M\\
\end{array}\right\rgroup
\eeq
where $D_x\equiv\partial_x-ieEt\,,$ whilst 
the related second order differential operator turns out to be
\beq
\left(i\partial\!\!\!\!/ + eA\!\!\!\!/+M\right)
\left(i\partial\!\!\!\!/ + eA\!\!\!\!/-M\right)
= P^{\,2}-M^2+e\sigma^{\,\mu\nu}F_{\mu\nu}
\label{qquadratic}
\eeq
where $\sigma^{\,\mu\nu}\equiv(i/4)[\,\gamma^\mu,\gamma^\nu\,]$
so that the 2nd order differential operator becomes
\beq
&&\left(i\partial\!\!\!\!/ + eA\!\!\!\!/+M\right)
\left(i\partial\!\!\!\!/ + eA\!\!\!\!/-M\right)
\equiv P^{\,2}-M^2+e\sigma^{\,\mu\nu}F_{\mu\nu}\nonumber\\
&& =\ -\partial_t^2-M^2+(\partial_x-ieEt)^2+\partial_y^2+\partial_z^2
+ ieE\,\alpha^1
\label{2ordine}
\eeq
It is convenient to obtain the solution of the Dirac equation
from the second order equation
\beq
\left(P^{\,2}-M^2+e\sigma^{\,\mu\nu}F_{\mu\nu}\right) f(x)\,\Upsilon=0
\label{ansatz}
\eeq
where $f(x)$ is a Lorentz invariant complex scalar function, while $\Upsilon$ is one of the
constant eigenbispinors of the matrix 
\beq
\fl
e\sigma^{\,\mu\nu}F_{\mu\nu}=
{ieE}\,\alpha^1=
{ieE}
\left\lgroup\begin{array}{cc}
  0& \sigma_x\\
     \sigma_x  & 0\\
\end{array}\right\rgroup\ =\
{ieE}
\left\lgroup\begin{array}{cccc}
0& 0& 0& 1\\
0& 0& 1& 0\\
0& 1& 0& 0\\
1& 0& 0& 0\\
\end{array}\right\rgroup
\eeq
Since the above matrix commutes with the matrix of the spin component
along the direction of the electrostatic field, i.e.,
\beq
[\,\alpha^1\,,\,\Sigma_1\,]=0\qquad\quad
\Sigma_1\equiv\textstyle\frac12\,i\,\gamma^2\gamma^3
=\frac12\left\lgroup\begin{array}{cc}
\sigma_x & 0\\
0 & \sigma_x\\
\end{array}\right\rgroup
\eeq
we can suitably introduce the four real bispinors
\beq
\fl
\Upsilon_{+}^{\,\uparrow}=
\textstyle\frac12\left\lgroup
\begin{array}{cccc}
1\\
1\\
1\\
1\\
\end{array}\right\rgroup\quad
\Upsilon_{+}^{\,\downarrow}=
\textstyle\frac12\left\lgroup
\begin{array}{cccc}
1\\
-1\\
-1\\
1\\
\end{array}\right\rgroup\qquad
\Upsilon_{-}^{\,\uparrow}=
\textstyle\frac12\left\lgroup\begin{array}{cccc}
1\\
1\\
-1\\
-1\\
\end{array}\right\rgroup\quad
\Upsilon_{-}^{\,\downarrow}=
\textstyle\frac12\left\lgroup\begin{array}{cccc}
1\\
-1\\
1\\
-1\\
\end{array}\right\rgroup
\eeq
\beq
\fl
e\sigma^{\,\mu\nu}F_{\mu\nu}\Upsilon_{\pm}^{\,r}
=\pm\,ieE\,\Upsilon_{\pm}^{\,r}
\qquad\quad
\beta\,\Upsilon_{\pm}^{\,r}=\Upsilon_{\mp}^{\,r}
\qquad\quad
(\,r\,=\,\uparrow,\downarrow\,)
\eeq
which satisfy the orthonormality and completeness relations
\beq
\alpha^1\,\Upsilon_{\pm}^{\,r}=\pm\,\Upsilon_{\pm}^{\,r}
\qquad\quad
\bar\Upsilon_{\pm}^{\,r}\,\beta\,\Upsilon_{\pm}^{\,s}=\delta^{\,rs}
\qquad\quad
(\,r,s=\uparrow\downarrow\,)\\
\fl\qquad
\Sigma_1\,\Upsilon_{\pm}^{\,\uparrow}=\Upsilon_{\pm}^{\,\uparrow}\qquad
\Sigma_1\,\Upsilon_{\pm}^{\,\downarrow}=\;-\,\Upsilon_{\pm}^{\,\downarrow}
\qquad\bar\Upsilon_{\pm}^{\,r}\,\beta\,\Upsilon_{\mp}^{\,s}=0
\qquad\quad
(\,r,s=\uparrow\downarrow\,)
\eeq
On the one hand, the bispinors of the first couple 
$\{\Psi_{+}^{\,r}(x)\,|\,r=\uparrow\downarrow\}\,,$
corresponding to the eigenvalue $+ieE\,,$ 
are mutually orthogonal and can be determined from the relationship
\beq
\Psi_{+}^{\,r}(x)=\left(i\partial\!\!\!/ + eA\!\!\!\!/+M\right)\,
f(x)\,\Upsilon_{+}^{\,r}\qquad\quad (\,r\,=\,\uparrow\downarrow\,)
\label{solutions12}
\eeq
On the other hand, the bispinors of the second couple 
$\{\Psi_{-}^{\,r}(x)\,|\,r=\uparrow\downarrow\}\,,$
related to the eigenvalue $-ieE\,,$ 
are in turn mutually orthogonal and are constructed according to
\beq
\Psi_{-}^{\,r}(x)&=&\left(i\partial\!\!\!/ + eA\!\!\!\!/+M\right)\,
f^\ast(x)\,\Upsilon_{-}^{\,r}\no
&=&\left(i\partial\!\!\!/ + eA\!\!\!\!/+M\right)\,
\beta\,\Upsilon_{+}^{\,r}\,f^\ast(x)
\qquad\quad (\,r\,=\,\uparrow\downarrow\,)
\label{solutions34}
\eeq
From eq.~(\ref{ansatz}) we obtain the 
second order differential equation for the eigenvalue $+ieE$ and the scalar functions
$f(x)$ which reads
\beq
&&\left[\,\partial_t^2+M^2-(\partial_x-ieEt)^2-\partial_y^2-\partial_z^2
- ieE\,\right]\,f(x)=0
\eeq
so that, after turning to the momentum space
\beq
f(x)&=&f(t,{\bf r})\equiv (2\pi)^{-3/2}\int \rmd{\bf p}\ 
\exp\{i\,{\bf p}\cdot{\bf r}\}\,\widetilde f(t,{\bf p})
\nonumber\\
&=& (2\pi)^{-3/2}\int \rmd{\bf p}\ 
\exp\{i\,p^kx^k\}\,\widetilde f(t,p^1,p^2,p^3)
\eeq
we eventually obtain, with $(p_x,p_y,p_z)\equiv (p^1,p^2,p^3)\,,$
\beq
\left[\,\partial_t^2+M^2+(p_x-eEt)^2+p_y^2+p_z^2-ieE\,\right]
\widetilde f(t,{\bf p})=0
\label{equadiffe}
\eeq
It is convenient to introduce the dimensionless quantities
\beq
\xi\equiv\frac{p_x-eEt}{\sqrt{eE}}\,,\qquad
\lambda\equiv\frac{p_y^2+p_z^2+M^2}{eE}
\eeq
so that we can rewrite the above equation in the standard form
\beq
\left(\partial_\xi^2+\xi^2+\lambda-i\right)\,\widetilde f(\xi,\lambda)=0
\nonumber
\eeq

It can be verified (see the appendix) that the forthcoming two
couples of linearly independent solutions of the above equation 
actually occur and read
\beq
\widetilde f_\lambda^{\,(1)}(\,\pm\,z_-\,)=D_{\,i\lambda/2}\,[\,\pm(1-i)\,\xi\,]
\label{f_1}\\
\widetilde f_\lambda^{\,(2)}(\,\pm\,z_+\,)=D_{\,-i\lambda/2-1}\,[\,\pm(1+i)\,\xi\,]
\label{f_2}\\
z_\pm\equiv(1\pm i)\,\xi=z_\mp^\ast
\eeq
where $D_\nu(z)$ are the parabolic cylinder functions.

In order to construct
the true physically relevant solutions of the first order Dirac
problem, we have to take carefully into account 
the physical content of the parabolic cylinder
functions, according to their asymptotic behaviour \cite{gradshteyn}
-- see the Appendix.
Namely, to a given particle of momentum ${\bf p}$ 
we shall associate the moving wave fronts, i.e.,
the surfaces of stationary phases
\[
\fl\qquad
\exp\{i\phi_\pm(t,{\bf r})\}\equiv\exp\{i{\bf p}\cdot{\bf r}\pm{\textstyle\frac12}\,i{\xi^2}(t)\}=\rm constant
\]
Then, in so doing, the temporal evolution is provided by
\[
\fl\qquad
i\partial_t\,\exp\{i\phi_\pm(t,{\bf r})\}=-\,\dot\phi_\pm(t,{\bf r})\,\exp\{i\phi_\pm(t,{\bf r})\}
=\mp\,\xi(t)\,\sqrt{eE}\,\exp\{i\phi_\pm(t,{\bf r})\}
\]
in such a manner that the phase factors $\phi_\pm(t,{\bf r})$ lead to positive frequency
solutions when $t\,\to\,\mp\,\infty$ that will describe, according to the free field theory jargon,
the particles of given momentum $\bf p$ and negative charge $(-\,e)\,.$

The normalized solutions of the original first order Dirac equation can
be obtained either from eq.~(\ref{solutions12}) or from eq.~(\ref{solutions34}),
the final achievement being the same.
The result is that we can definitely build up two complete 
orthonormal sets of {\sl in}--states: namely,
$u_{{\bf p},\,r}^{(-)}\,(t,{\bf r})\,,$ describing 
incoming electrons of momentum ${\bf p}$ and helicity $r$ 
while $v_{{\bf p},\,r}^{(-)}\,(t,{\bf r})$ 
will describe incoming positrons of momentum ${\bf p}$ and helicity $r$ 
respectively.
Analogously, we can in turn construct the complete 
orthonormal sets of {\sl out}--states\ :  $u_{{\bf p},\,r}^{(+)}\,(t,{\bf r})\,,$ describing
outgoing electrons of momentum ${\bf p}$ and helicity $r$ and $v_{{\bf p},\,r}^{(+)}\,(t,{\bf r})\,,$
which describe outgoing positrons of momentum ${\bf p}$ and helicity $r$ respectively.
Taking all of that into account we have 
\beq
\fl\qquad
&& \left(\,\gamma^{\,\mu}P_\mu\ +\ M\,\right) \exp\{i\,{\bf p}\cdot{\bf r}\}=
\nonumber\\
\fl\qquad
&& \left\lgroup\begin{array}{cccc}
-\sqrt{eE}\,i\rmd_\xi + M& 0& -p_z& -\xi\sqrt{eE}+ip_y\\
0& -\sqrt{eE}\,i\rmd_\xi + M& -\xi\sqrt{eE}-ip_y& p_z\\
p_z& \xi\sqrt{eE}-ip_y& \sqrt{eE}\,i\rmd_\xi + M& 0\\
\xi\sqrt{eE}+ip_y& -p_z& 0& \sqrt{eE}\,i\rmd_\xi + M\\
\end{array}\right\rgroup{\rm e}^{\,i\,{\bf p}\cdot{\bf r}}
\nonumber
\eeq
after setting
\[
i\partial_t=-\sqrt{eE}\,i\rmd_\xi=-\sqrt{eE}\,(1\pm i)\frac{i\rmd}{\rmd z_\pm}
\]
we immediately obtain
\beq
(\gamma^{\,\mu}P_\mu + M)\,\Upsilon_{+}^{\,\uparrow} =
M\,\Upsilon_{+}^{\,\uparrow}(\varpi) -
\Upsilon_{-}^{\,\uparrow}\sqrt{eE}\,(\xi+i\rmd_\xi)\\
\Upsilon_{+}^{\,\uparrow}(\varpi)\equiv
\Upsilon_{+}^{\,\uparrow} + i\,\varpi\,\Upsilon_{+}^{\,\downarrow}
\qquad\qquad
\varpi\equiv(\,{p_y+ip_z}\,)/{M}
\label{polarization1}
\eeq
and from the recursive relations
\beq
(\,\xi+i\rmd_\xi\,)\,D_{\,i\lambda/2}\,(\pm\,z_-)=
\mp\,{\textstyle\frac12}\,{\lambda}\,(1-i)\,D_{\,i\lambda/2-1}\,(\pm\,z_-)
\eeq
we eventually find
\beq
&& (\gamma^\mu P_\mu + M)\,\Upsilon_{+}^{\,\uparrow}\,D_{\,i\lambda/2}\,(\pm\,z_-)=
\nonumber\\
&& \Upsilon_{+}^{\,\uparrow}(\varpi)\,MD_{\,i\lambda/2}\,(\pm\,z_-) \pm
{\textstyle\frac12}\,\lambda\,(1-i)\,\sqrt{eE}\,
D_{\,i\lambda/2-1}\,(\pm\,z_-)\,\Upsilon_{-}^{\,\uparrow} 
\nonumber
\eeq
In a quite analogous way we obtain
\beq
(\gamma^\mu P_\mu + M)\,\Upsilon_{+}^{\,\downarrow} =
M\,\Upsilon_{+}^{\,\downarrow}(\varpi) -
\Upsilon_{-}^{\,\downarrow}\sqrt{eE}\,(\xi+i\rmd_\xi)\\
\Upsilon_{+}^{\,\downarrow}(\varpi)\equiv
\Upsilon_{+}^{\,\downarrow} + i\,\varpi\,\Upsilon_{+}^{\,\uparrow}
\label{polarization2}
\eeq
and consequently
\beq
&& (\gamma^\mu P_\mu + M)\,\Upsilon_{+}^{\,\downarrow}\,D_{\,i\lambda/2}\,(\pm\,z_-)=
\nonumber\\
&& \Upsilon_{+}^{\,\downarrow}(\varpi)\,MD_{\,i\lambda/2}\,(\pm\,z_-)\pm
{\textstyle\frac12}\,\lambda\,(1-i)\,\sqrt{eE}\,
D_{\,i\lambda/2-1}\,(\pm\,z_-)\,\Upsilon_{-}^{\,\downarrow} 
\nonumber
\eeq
It follows therefrom that the bispinor solutions which describe an 
incoming electron can be written as\quad$(\,r=\,\uparrow,\downarrow\,)$
\beq
&& u_{\,{\bf p},\,r}^{\,\rm in}\,(t,{\bf r})\ =\ [\,\lambda eE(2\pi)^3\,]^{-1/2}\,
\exp\left\{i\,{\bf p}\cdot{\bf r} -\,{\textstyle\frac18}{\pi\lambda}\right\}
\nonumber\\
&& \times\ \left\lbrace
\Upsilon_{+}^{\,r}(\varpi)\,MD_{\,i\lambda/2}\,(z_-) +
{\textstyle\frac12}\,\lambda\,(1-i)\,\sqrt{eE}\,D_{\,i\lambda/2-1}\,(z_-)\,
\Upsilon_{-}^{\,r}
\right\rbrace
\nonumber
\eeq
whilst the bispinor solutions which describe an outgoing positron read
\beq
&& v_{\,{\bf p},\,r}^{\,\rm out}\,(t,{\bf r})\ =\ [\,\lambda eE(2\pi)^3\,]^{-1/2}\,
\exp\left\{i\,{\bf p}\cdot{\bf r} -\,{\textstyle\frac18}{\pi\lambda}\right\}
\nonumber\\
&& \times\ \left\lbrace \Upsilon_{+}^{\,r}(\varpi)\,MD_{\,i\lambda/2}\,(-\,z_-) -
{\textstyle\frac12}\,\lambda\,(1-i)\,\sqrt{eE}\,D_{\,i\lambda/2-1}\,(-\,z_-)\,
\Upsilon_{-}^{\,r}
\right\rbrace
\nonumber
\eeq
Notice that we actually obtain the 
standard normalization\quad$(\,r,s=\,\uparrow,\downarrow\,)$
\beq
\fl\quad
\int \rmd{\bf r}\ \bar u_{\,{\bf p},\,r}^{\,\rm in}\,(t,{\bf r})\,\beta\,
u_{\,{\bf q}\,,\,s}^{\,\rm in}\,(t,{\bf r}) =
\delta({\bf p}-{\bf q})\,\delta_{rs}
= \int \rmd{\bf r}\ \bar v_{\,{\bf p}\,,\,r}^{\,\rm out}\,(t,{\bf r})\,\beta\,
v_{\,{\bf q}\,,\,s}^{\,\rm out}\,(t,{\bf r}) 
\eeq

A closely similar construction can be done to set up the bispinor
solutions which describe incoming positrons and outgoing electrons  respectively.
As a matter of fact, from the other linearly independent solutions (\ref{f_2})
we get
\beq
(\gamma^\mu P_\mu + M)\,\Upsilon_{+}^{\,r}\,D_{\,-i\lambda/2-1}\,(\pm\,z_+)=\no
\Upsilon_{+}^{\,r}(\varpi)\,MD_{\,-i\lambda/2-1}\,(\pm\,z_+)\mp
\sqrt{eE}\,D_{\,-i\lambda/2}\,(\pm\,z_+)\,\Upsilon_{-}^{\,r}
\eeq
where use has been made of the recursive relation
\[
(\,\xi+i\rmd_\xi\,)\,D_{\,-i\lambda/2-1}\,(\pm\,z_+)=\pm\,(1-i)\,D_{\,-i\lambda/2}\,(\pm\,z_+)
\]
so that we eventually obtain the following list of
incoming and outgoing bispinor solutions of the Dirac equation
(\ref{eeeqdirac}) with a given momentum
and polarization
\beq
&& u_{\,{\bf p},\,r}^{\,\rm in}\,(t,{\bf r})\ =\ [\,\lambda eE(2\pi)^3\,]^{-1/2}\,
\exp\left\{i\,{\bf p}\cdot{\bf r} -\,{\textstyle\frac18}{\pi\lambda}\right\}
\nonumber\\
&& \times\ \left\lbrace
\Upsilon_{+}^{\,r}(\varpi)\,MD_{\,i\lambda/2}\,(z_-) +
{\textstyle\frac12}\,\lambda\,(1-i)\,\sqrt{eE}\,D_{\,i\lambda/2-1}\,(z_-)\,
\Upsilon_{-}^{\,r}
\right\rbrace
\label{inelectron_r}
\eeq
\beq
&& v_{\,{\bf p}\,,\,s}^{\,\rm in}\,(t,{\bf r})\ =\ [\,2eE\,(2\pi)^3\,]^{-1/2}\,
\exp\left\{i\,{\bf p}\cdot{\bf r} -\,{\textstyle\frac18}{\pi\lambda}\right\}\no
&&\times \left\lbrace\Upsilon_{+}^{\,s}(\varpi)\,M
D_{-i\lambda/2-1}\,(z_+) -
\,(1-i)\,\sqrt{eE}\;D_{-i\lambda/2}\,(z_+)\,
\Upsilon_{-}^{\,s}
\right\rbrace
\label{inpositron_s}
\eeq
\beq
\fl\qquad\quad
&& u_{\,{\bf p}\,,\,s}^{\,\rm out}\,(t,{\bf r})\ =\ [\,2eE\,(2\pi)^3\,]^{-1/2}\,
\exp\left\{i\,{\bf p}\cdot{\bf r} -\,{\textstyle\frac18}{\pi\lambda}\right\}\no
\fl\qquad\quad
&&\times \left\lbrace\Upsilon_{+}^{\,s}(\varpi)\,M
D_{-i\lambda/2-1}\,(-\,z_+) +
\,(1-i)\,\sqrt{eE}\;D_{-i\lambda/2}\,(-\,z_+)\,
\Upsilon_{-}^{\,s}
\right\rbrace
\label{outelectron_s}
\eeq
\beq
\fl\qquad\quad
&& v_{\,{\bf p},\,r}^{\,\rm out}\,(t,{\bf r})\ =\ [\,\lambda eE(2\pi)^3\,]^{-1/2}\,
\exp\left\{i\,{\bf p}\cdot{\bf r} -\,{\textstyle\frac18}{\pi\lambda}\right\}
\nonumber\\
\fl\qquad\quad
&& \times\ \left\lbrace \Upsilon_{+}^{\,r}(\varpi)\,MD_{\,i\lambda/2}\,(-\,z_-) -
{\textstyle\frac12}\,\lambda\,(1-i)\,\sqrt{eE}\,D_{\,i\lambda/2-1}\,(-\,z_-)\,
\Upsilon_{-}^{\,r}
\right\rbrace
\label{outpositron_r}
\eeq
By looking at the asymptotic behaviour of the parabolic cylinder functions
-- see the appendix -- one obtains the following result, in accordance with the
previously discussed meaning of the positive and negative frequencies solutions:
namely,
\beq
\fl\quad
u_{\,{\bf p}\,,\,s}^{\,\rm in}\,(t,{\bf r})\sim\;
\frac{M\,[\,2\xi^2(t)\,]^{\,i\lambda/4}}
{\surd\,[\,\lambda eE(2\pi)^3\,]}
\exp\left\{i\,{\bf p}\cdot{\bf r}+{\textstyle\frac12}\,i\,\xi^2(t)\right\}
\Upsilon_{+}^{\,s}(\varpi)\qquad
(\,t\to-\,\infty\,)\\
\fl\quad
v_{\,{\bf p}\,,\,s}^{\,\rm in}\,(t,{\bf r})\sim\;
\frac{[\,2\xi^2(t)\,]^{\,-\,i\lambda/4}}{\surd\,[\,2eE(2\pi)^3\,]}
\exp\left\{i\,{\bf p}\cdot{\bf r}-{\textstyle\frac12}\,i\,\xi^2(t)
+{\textstyle\frac34}\,\pi i\right\}\,
\Upsilon_{-}^{\,s}\,\quad\;
(\,t\to-\,\infty\,)\\
\fl\quad
u_{\,{\bf p}\,,\,s}^{\,\rm out}\,(t,{\bf r})\sim\;
\frac{[\,2\xi^2(t)\,]^{\,-\,i\lambda/4}}{\surd\,[\,2eE(2\pi)^3\,]}\,
\exp\left\{i\,{\bf p}\cdot{\bf r}-{\textstyle\frac12}\,i\,\xi^2(t)
- {\textstyle\frac14}\,\pi i\right\}\,
\Upsilon_{-}^{\,s}\quad\;
(\,t\to +\,\infty\,)\\
\fl\quad
v_{\,{\bf p}\,,\,s}^{\,\rm out}\,(t,{\bf r})\sim\;
\frac{M\,[\,2\xi^2(t)\,]^{\,i\lambda/4}}{\surd\,[\,\lambda eE(2\pi)^3\,]}\,
\exp\left\{i\,{\bf p}\cdot{\bf r}+{\textstyle\frac12}\,i\,\xi^2(t)\right\}
\Upsilon_{+}^{\,s}(\varpi)\qquad
(\,t\to +\,\infty\,)
\eeq

\medskip
The two sets of incoming and outgoing bispinor wave functions
of definite momentum and polarization 
\beq
\{u_{\,{\bf p}\,,\,r}^{\,\rm in}\,;\,v_{\,{\bf p}\,,\,r}^{\,\rm in}\}\,,
\{u_{\,{\bf q}\,,\,s}^{\,\rm out}\,;\,v_{\,{\bf q}\,,\,s}^{\,\rm out}\}
\qquad\quad(\,{\mathbf p},{\mathbf q}\in{\mathbb R}^3\,,\ r,s=\,\uparrow,\downarrow\,)
\label{sets}
\eeq
turns out to be orthonormal and complete in the generalized sense, i.e.
in the sense of the theory of the tempered distributions.
Thus they represent the bases in the Hilbert spaces
${\mathfrak H}^{\,\rm in}$ and ${\mathfrak H}^{\,\rm out}$
of the Dirac bispinors in the presence of an electrostatic
background field.
As a matter of fact, we can readily verify that
the scalar products are time independent owing to 
the self--adjointness of the 1-particle Dirac hamiltonian 
$H_t$ so that
\beq
\left(\,u_{\,{\bf q}\,,\,r}^{\,\rm in}\,,\,u_{\,{\bf p}\,,\,s}^{\,\rm in}\,\right)
&\equiv& \int \rmd{\mathbf r}\
\bar u_{\,{\bf q}\,,\,r}^{\,\rm in}(t,{\bf r})\,\beta\,
u_{\,{\bf p}\,,\,s}^{\,\rm in}(t,{\bf r})\no
&=&\delta ({\bf p}-{\bf q})\,\delta_{\,rs}
=\left(\,u_{\,{\bf q}\,,\,r}^{\,\rm out}\,,\,u_{\,{\bf p}\,,\,s}^{\,\rm out}\,\right)\\
\left(\,v_{\,{\bf q}\,,\,r}^{\,\rm in}\,,\,v_{\,{\bf p}\,,\,s}^{\,\rm in}\,\right)
&\equiv& \int \rmd{\mathbf r}\
\bar v_{\,{\bf q}\,,\,r}^{\,\rm in}(t,{\bf r})\,\beta\,
v_{\,{\bf p}\,,\,s}^{\,\rm in}(t,{\bf r})\no
&=&\delta ({\bf p}-{\bf q})\,\delta_{\,rs}
=\left(\,v_{\,{\bf q}\,,\,r}^{\,\rm out}\,,\,v_{\,{\bf p}\,,\,s}^{\,\rm out}\,\right)
\eeq
Notice that we can also immediately obtain that
\beq
\fl\qquad
\left(\,u_{\,{\bf q}\,,\,s}^{\,\rm in}\,,\,v_{\,{\bf p}\,,\,r}^{\,\rm in}\,\right)
\equiv\int \rmd{\mathbf r}\
\bar u_{\,{\bf q}\,,\,s}^{\,\rm in}(t,{\bf r})\,\beta\,
v_{\,{\bf p}\,,\,r}^{\,\rm in}(t,{\bf r})=0\\
\fl\qquad
\left(\,v_{\,{\bf q}\,,\,s}^{\,\rm out}\,,\,u_{\,{\bf p}\,,\,r}^{\,\rm out}\,\right)
\equiv\int \rmd{\mathbf r}\
\bar v_{\,{\bf q}\,,\,s}^{\,\rm out}(t,{\bf r})\,\beta\,
u_{\,{\bf p}\,,\,r}^{\,\rm out}(t,{\bf r})=0
\eeq
which means that the positive and negative frequency
solutions of the Dirac equations are mutually orthogonal. 
Moreover, it is straightforward although tedious to check 
by direct inspection
the equal time closure relations
\beq
&& \sum_{{\bf p}\,,\,r}\left[\,
u_{\,{\bf p}\,,\,r}^{\,\rm as}\,\otimes\,
\bar u_{\,{\bf p}\,,\,r}^{\,\rm as} +
v_{\,{\bf p}\,,\,r}^{\,\rm as}\,\otimes\,
\bar v_{\,{\bf p}\,,\,r}^{\,\rm as}\,\right]
= \beta
\label{completeness3}
\eeq
where we have set for the sake of brevity
\[
\sum_{{\bf p}\,,\,r}\ \equiv\ \int \rmd{\mathbf p}\sum_{r\,=\,\uparrow\downarrow}
\]
while $^{\,\rm as}$ stands for either $^{\,\rm in}$
or $^{\,\rm out}\,.$
As a consequence, we can write in the very same way the most general operator solutions 
of the Dirac equation as follows: namely,
\beq
\psi(x) &=& \sum_{{\bf p}\,,\,r}\
\left[\,a_{\;\!\rm as}({\bf p},r)\,u_{\,{\bf p}\,,\,r}^{\,\rm as}\,(x)+
b_{\rm as}^{\,\dagger}({\bf p},r)\,v_{\,{\bf p}\,,\,r}^{\,\rm as}\,(x)\,\right]
\\
{\bar\psi}(x) &=& \sum_{{\bf p}\,,\,r}\
\left[\,a^{\,\dagger}_{\;\!\rm as}({\bf p},r)\,\bar u_{\,{\bf p}\,,\,r}^{\,\rm as}\,(x)+
b_{\rm as}({\bf p},r)\,\bar v_{\,{\bf p}\,,\,r}^{\,\rm as}\,(x)\,\right]
\eeq
where the creation and destruction operators satisfy the
canonical anticommutation relations
\beq
&& \{a_{\;\!\rm as}({\bf p},r),a_{\;\!\rm as}^{\,\dagger}({\bf q},s)\}=
\delta({\bf p}-{\bf q})\,\delta_{\,rs} =
\{b_{\rm as}({\bf p},r),b_{\rm as}^{\,\dagger}({\bf q},s)\}
\nonumber
\eeq
all the other anticommutators being equal to zero. From the closure relations
(\ref{completeness3}) we can easily derive the canonical equal time 
anticommutation relations
\beq
\fl
\{\psi(t,{\bf r}'),\bar\psi(t,{\bf r})\} =
\sum_{{\bf p}\,,\,r}\ \left[\,u_{\,{\bf p}\,,\,r}^{\,\rm as}\,(t,{\bf r}')\,
\bar u_{\,{\bf p}\,,\,r}^{\,\rm as}\,(t,{\bf r})
+ v_{\,{\bf p}\,,\,r}^{\,\rm as}\,(t,{\bf r}')\,
\bar v_{\,{\bf p}\,,\,r}^{\,\rm as}\,(t,{\bf r})\,\right]
= \beta\;\delta({\bf r}-{\bf r}')
\nonumber
\eeq
Accordingly, the causal Green's function, or Feynman propagator, for the
Dirac spinor under the influence of a background electrostatic
homogeneous field has been firstly reported in Ref.~\cite{nikishov}
in full detail.

This means that we have to introduce the Fock spaces 
$\mathfrak F^{\,\rm as}$ for the
incoming and outgoing particles and antiparticles.
For example, starting from the in--vacuum $|\,0\,\rm in\,\rangle$
one can generate as usual the 1-particle state of charge $-\,e$ momentum $\bf p$
and polarization $s=\uparrow\downarrow$ describing an incoming electron, viz.,
\[
a_{\;\!\rm in}^{\,\dagger}({\bf p},s)|\,0\,{\rm in}\,\rangle
=|\,-\,{\bf p}\,s\;{\rm in}\,\rangle
\]
the corresponding wave function being
$u_{\,{\bf p}\,,\,s}^{\,\rm in}(t,{\bf r})\,.$
All the other 1-particle and many particle states for incoming and outgoing
electrons and positrons can be constructed in close analogy.
It follows therefrom that the charge operator, the
momentum operator and the helicity operator, i.e. the projection of
the spin angular momentum along the electric field direction, 
will correspond to the customary expressions
\beq
Q&\equiv&(-\,e)\int \rmd{\mathbf r}\
:{\psi}^\dagger(t,{\bf r})\,\psi(t,{\bf r}):
\nonumber\\
&=& (-\,e)\sum_{{\bf p}\,,\,r}\
\left[\,a^{\dagger}({\bf p}\,,r)\,a\,({\bf p}\,,r)
- b^{\,\dagger}({\bf p}\,,r)\,b\,({\bf p}\,,r)\,\right]\\
{\bf P} &\equiv& (-\,i)\int \rmd{\mathbf r}\
:{\psi}^\dagger(t,{\bf r})\,\nabla\,\psi(t,{\bf r}):
\nonumber\\
&=& \sum_{{\bf p}\,,\,r}\ {\bf p}\,
\left[\,a^{\dagger}({\bf p}\,,r)\,a\,({\bf p}\,,r)
- b^{\,\dagger}({\bf p}\,,r)\,b\,({\bf p}\,,r)\,\right]\\
h &\equiv& {\textstyle\frac12}\int_{-\infty}^\infty\rmd x
:{\psi}^\dagger(t,x)\,\Sigma_1\,\psi(t,x):\nonumber\\
&=&{\textstyle\frac12}\int_{-\infty}^\infty\rmd p\
\Big[\,a^{\dagger}(\,{p}\,,\uparrow)\,a\,(\,p\,,\uparrow) - 
a^{\dagger}(\,{p}\,,\downarrow)\,a\,(\,p\,,\downarrow)\nonumber\\
&+&  b^{\dagger}(\,{p}\,,\uparrow)\,b\,(\,p\,,\uparrow) - 
b^{\dagger}(\,{p}\,,\downarrow)\,b\,(\,p\,,\downarrow)\,\Big]
\eeq
where the suffix $^{\rm in}$ or $^{\rm out}$ is understood.

\bigskip
Let us now turn to the calculation of the invariant inner product between the wave functions of an
incoming electron with quantum numbers ${\bf p}\,,\,r$ and of an outgoing
positron of quantum numbers ${\bf q}\,,\,s\,:$ we find
\beq
\int\rmd{\mathbf r}\
\bar v_{\,{\bf q}\,,\,s}^{\,\rm out}\,(t,{\bf r})\,\beta\,u_{\,{\bf p}\,,\,r}^{\,\rm in}\,(t,{\bf r})
=\e^{-\,\pi\lambda/2}\,\delta({\bf p}-{\bf q})\,\delta_{\,rs}
\eeq
and analogously
\beq
\int\rmd{\mathbf r}\
\bar v_{\,{\bf q}\,,\,s}^{\,\rm in}\,(t,{\bf r})\,\beta\,u_{\,{\bf p}\,,\,r}^{\,\rm out}\,(t,{\bf r})
=e^{-\,\pi\lambda/2}\,
\delta({\bf p}-{\bf q})\,\delta_{\,rs}
\eeq
so that we can write in a compact way
\beq
\left(\,v_{\,{\bf q}\,,\,s}^{\,\rm in}\,,\,u_{\,{\bf p}\,,\,r}^{\,\rm out}\,\right)
= e^{-\,\pi\lambda/2}\,
\delta({\bf p}-{\bf q})\,\delta_{\,rs}
=\left(\,u_{\,{\bf q}\,,\,s}^{\,\rm in}\,,\,v_{\,{\bf p}\,,\,r}^{\,\rm out}\,\right)\\
\forall\,{\bf p},{\bf q}\in{\mathbb R}^3\,,\ r,s=\uparrow\downarrow
\nonumber
\eeq
The above equality has led Nikishov \cite{nikishov} to the attempt of understanding
the real positive quantity
\[
w_\lambda\equiv\exp\{-\,\pi\,(\,p_y^2+p_z^2+M^2\,)/eE\}
\qquad\quad0<w_\lambda<1
\]
as the probability of creating out of the vacuum one electron positron pair,
in which the particle and the antiparticle have opposite good quantum numbers
$(\,\mp\,e\,,{\bf p},r\,)\,.$ This intringuing interpretation is supported by the fact
that indeed we have
\beq
\left(\,u_{\,{\bf q}\,,\,s}^{\,\rm out}\,,\,u_{\,{\bf p}\,,\,r}^{\,\rm in}\,\right)\
=\  \delta_{\,rs}\,\delta({\bf p}-{\bf q})\;N_\lambda
\eeq
where we have set
\beq
N_\lambda\ \equiv\ \exp\left\{-\,{\pi\lambda/4}\right\}\,\sqrt{{\lambda\over\pi}}\;
\Gamma\left({i\lambda\over 2}\right)\,\sinh{\pi\lambda\over 2}
\label{N_lambda}
\eeq
Now, since we find
\[
N_\lambda\,N_\lambda^\ast=1-\e^{-\,\pi\lambda}=1-w_\lambda
\]
it is natural to understand the complex quantity $N_\lambda$
as the probability amplitude that one pair, in which the particle and the antiparticle 
have opposite quantum numbers
$(\,\mp\,e\,,{\bf p},r\,)\,,$ is {\bf not} created out of the vacuum by
absorbing energy from the electrostatic external field.
Actually, in fact, it is also straightforward to verify that for positrons we get
\beq
\int\rmd{\mathbf r}\
\bar v_{\,{\bf q}\,,\,s}^{\,\rm out}\,(t,{\bf r})\,\beta\,v_{\,{\bf p}\,,\,r}^{\,\rm in}\,(t,{\bf r})
=
N_\lambda^\ast\,\delta({\bf p}-{\bf q})\,\delta_{\,rs}
\eeq
thus providing the full endorsement to the whole construction.
By the way, it is apparent that the probability amplitudes
for the absorption and nonabsorption of one pair $(\,\mp\,e\,,{\bf p},r\,)\,,$
are still expressed by $w_\lambda$ and $1-w_\lambda$ 
respectively, owing to obvious symmetry reasons.
\section{Pairs Production Annihilation Mechanism}
Here I would like to shortly discuss how, according to the Nikishov's original proposal \cite{nikishov},
one can try to recover the Schwinger formula 
for the rate of pairs emission and absorption, per unit time and unit volume,
from the knowledge of the exact solutions of the Dirac equation. 
As we shall see here below, this attempt is not at all flawless,
because of the presence of divergences and the consequent unavoidable
introduction of some regularisation method.
To this concern, it is useful to
perform the Bogolyubov transformations connecting the incoming and outgoing
complete orthonormal sets of the exact nonperturbative solutions of the Dirac equation 
in the presence of the homogeneous electrostatic field.

The above four sets (\ref{sets}) of incoming and outgoing solutions are indeed related
throughout a Bogolyubov  transformation, that is
\beq
u_{\,{\bf p}\,,\,r}^{\,\rm in} = c_{\,1\lambda}\,u_{\,{\bf p}\,,\,r}^{\,\rm out} + 
c_{\,2\lambda}\;v_{\,{\bf p}\,,\,r}^{\,\rm out}\\
v_{\,{\bf p}\,,\,r}^{\,\rm in} = c_{\,2\lambda}^{\,\ast}\;u_{\,{\bf p}\,,\,r}^{\,\rm out} +
c_{\,1\lambda}^{*}\;v_{\,{\bf p}\,,\,r}^{\,\rm out}
\\
|c_{\,1\lambda}|^2+|c_{\,2\lambda}|^2=1
\label{bogolyubov}
\eeq
We have
\beq
\left(\,u_{\,{\bf p}\,,\,r}^{\,\rm out}\,,\,u_{\,{\bf q}\,,\,s}^{\,\rm in}\,\right)=
N_\lambda\,\delta({\bf p}-{\bf q})\,\delta_{\,rs}\\
\left(\,v_{\,{\bf p}\,,\,r}^{\,\rm out}\,,\,v_{\,{\bf q}\,,\,s}^{\,\rm in}\,\right)=
N_\lambda^{*}\,\delta({\bf p}-{\bf q})\,\delta_{\,rs}\\
\fl\quad
\left(\,u_{\,{\bf p}\,,\,r}^{\,\rm out}\,,\,v_{\,{\bf q}\,,\,s}^{\,\rm in}\,\right)=
\exp\left\{-\,\textstyle\frac14\,{\pi\lambda}\right\}\,\delta({\bf p}-{\bf q})\,\delta_{\,rs}=
\left(\,u_{\,{\bf p}\,,\,r}^{\,\rm in}\,,\,v_{\,{\bf q}\,,\,s}^{\,\rm out}\,\right)
\eeq
Then, using the above listed orthonormality relations,
we immediately find
\beq
c_{\,1\lambda}=N_\lambda\qquad\quad 
c_{\,2\lambda}=\exp\left\{-\,\textstyle\frac14\,{\pi\lambda}\right\}=c_{\,2\lambda}^{\,\ast}
\eeq
In turn, the relative probability of a pair production
will be coherently given by
$$
w_{\lambda}=1-|\,N_\lambda\,|^2=c_{\,2\lambda}^{\,2}=\exp\left\{-\,{\pi\lambda}\right\}
$$

\bigskip\noindent
The vacuum to vacuum persistence amplitude can
be written in the form -- see also the recent up to date review
\cite{dunne},
\beq
\langle\,{\rm out}\,0\,|\,0\,{\rm in}\,\rangle=\exp\left\{{\rm
  i\over\hbar}\,[\,\R\,\Gamma_{\rm eff}(E) +{\rm i}\,\I\,\Gamma_{\rm eff}(E)\,]\right\}\\
|\,\langle\,{\rm out}\,0\,|\,0\,{\rm in}\,\rangle\,|^{\,2}=
\exp\{-\,2{\mathcal V}\,(\mathrm t_f-\mathrm t_{\,i})\,\Im{\rm m}\,{\cal L}_{\rm eff}(E)\}
\label{eff_Lagrange}
\eeq
where ${\cal L}_{\rm eff}(E)$ is referred to as the effective lagrangian density
in the presence of a background electrostatic field $E\,,$
whereas $\mathrm t_f-\mathrm t_{\,i}$ does indicate the very long total time 
during which the pairs production or annihilation takes place, while $\mathcal V$ denotes
the total volume of e.g. a very large cubic box of sides $L_x,L_y,L_z\,$
in the three dimensional space. Needless to say, the very large dimensional quantities
$\mathrm t_f-\mathrm t_{\,i}$ and $\mathcal V$ are in fact infrared divergences.
It turns out that $\Gamma_{\rm eff}(E)$ contains a real part that describes dispersive
effects like the Faraday's birifringence, as well as an imaginary part that
concerns absorbitive effects like vacuum pairs production.

Now, according to the above described interpretation, the
absolute  probability for no pairs creation will be thereof given by
the formal, divergent expression
\beq
|\,\langle\,{\rm out}\,0\,|\,0\,{\rm in}\,\rangle\,|^{\,2}
&=&\prod_{{\bf k}\,,\,r}\,N_\lambda\,N^\ast_\lambda
=\prod_{\bf k}\,\left(\,1-\e^{-\,\pi\lambda}\,\right)^2\no
&\equiv&
\exp\{\,2\textstyle\sum_{\bf k}\ln\left(\,1-\e^{-\,\pi\lambda}\,\right)\}
\label{vuoto_vuoto}
\eeq
where the mandatory dimensionless spatial vector index 
\[
{\bf k}\,=\,(\,p_x\,L_x/2\pi\,,\,p_y\,L_y/2\pi\,,\,p_z\,L_z/2\pi\,)
\]
has been introduced to vindicate the probabilistic interpretation with
\[
\sum_{\bf k}\,\equiv\,{\mathcal V}_n\,(2\pi)^{-n}\int\rmd{\bf p}
\]
Notice that when the electric field is switched off then $w_\lambda\to0$
and consequently the vacuum to vacuum probability becomes trivially
equal to one. It is important to realize that the validity of the above
relation (\ref{vuoto_vuoto}) stems from the natural assumption that
the production and the destruction of any two pairs with different
quantum numbers are statistically independent events.
Now we find
\beq
\ln\,|\,\langle\,{\rm out}\,0\,|\,0\,{\rm in}\,\rangle\,|^{\,2}
=2\,L_x\,L_y\,L_z\,(2\pi)^{-\,3}\int_{-\infty}^\infty\rmd p_x\no
\times\int_{-\infty}^\infty\rmd p_y\int_{-\infty}^\infty\rmd p_z\
\ln\left(\,1-\exp\{-\,\pi(\,p_y^2+p_z^2+M^2\,)/eE\}\,\right)
\eeq
which is, as it stands, an ill defined expression
owing to the infrared (large volume) and ultraviolet (large longitudinal
momentum $p_x$) divergences. The latter one can be expressed
in terms of the large total time during which the process of pairs emission or
absorption takes place: namely,
\[
\int_{-\infty}^\infty \rmd p_x\,\equiv\,eE(\mathrm t_f-\mathrm t_{\,i})
\]
what corresponds to a suitable ultraviolet regularisation because,
in the presence of a uniform field, this time interval is arbitrarily long.
To sum up, we can express the
total absolute probability of pairs production out of the vacuum
(or pairs absorption into the vacuum) as

\beq
W_{\,\rm pairs} &\equiv& 1-|\langle\,{\rm out}\,0\,|\,0\,{\rm
  in}\,\rangle|^{\,2}\no
&=& 1-\exp\left\{-\,{2
  \over\hbar}\,\I\,\Gamma_{\rm eff}(E)\right\}\no
&\approx& {2
  \over\hbar}\,{\mathcal V}\,(\mathrm t_f-\mathrm t_{\,i})\,\I\,{\cal L}_{\,\rm eff}(E)
\eeq
in which we have taken into account that the quantity $\I\,\Gamma_{\rm eff}(E)/\hbar$
is tipically very small for massive particle antiparticle pairs.
Hence, the average probability of
pairs production per unit volume and unit time is approximately given by
\beq
\Gamma_{\,\rm pairs} &\equiv& 
W_{\,\rm pairs}\,[\,{\mathcal V}\,(\mathrm t_f-\mathrm t_{\,i})\,]^{-1}
\;\approx\;
{2\over\hbar}\,\Im{\rm m}\,{\cal L}_{\rm eff}(E)\no
&=& -\,\frac{eE}{4\pi^3}
\int_{-\infty}^\infty \rmd p_y\int_{-\infty}^\infty \rmd p_z\ 
\ln\left(1-\exp\left\{-\,\pi\lambda\right\}\right)
\nonumber\\
&=& \frac{e^2E^2}{4\pi^3}\,\sum_{n=1}^\infty {1\over n^2}\,
\exp\left\{-\,n\,\frac{\pi M^2}{eE}\right\}
\label{instanton}
\eeq
the $n$th term of the series being {\em grosso modo}
understood to be the probability of the emission or absorpion of
$n$ pairs. 
From the above equation one can readily extract the value of the critical
electrostatic field, above which the probability of production
of e.g. electron positron pairs becomes appreciable: namely,
\[
E_{\,\rm cr}=\frac{m_e^2\,c^{\;\!3}}{\hbar\,e}\simeq 1.3\times10^{\,18}\ \rm V/m
\]
which is far beyond the present experimental capabilities.
Nonetheless it is worthwhile to observe to this concern, in accordance with
refs.~\cite{graphene}, that for massless charged spinor particle
the approximation $\frac12\,\Gamma_{\,\rm pairs}\approx 
\Im{\rm m}\,{\cal L}_{\rm eff}(E)$
does not certainly hold true. This opens the interesting possibility 
to detect the pairs production and destruction in a two dimensional
graphene sample.

As a final remark, I recall that, as it is well known, the nonperturbative complex effective action can be
rigorously derived from the euclidean formulation and the zeta function regularisation
-- see Appendix B.
\section{The Algebraic Approach}\label{algebraic}
In order to develop the most general algebraic approach to the pairs production and annihilation
processes in the presence of an external background uniform field, it is convenient
to introduce a multi-valued index $\imath,\jmath,\ell,\varkappa,\ldots$ to label
the whole set $\mathfrak Q$ of discrete and continuos quantum numbers that correspond to
the conserved quantities of the system under consideration,
but for the quantum number that distinguishes
particles from antiparticles.
For example, in the case of a spinor field in a constant homogeneous electric field 
on the four dimensional Minkowski spacetime we
have $\imath=(\,{\bf p},r\,)$ with ${\bf p}\in \mathbb R^3\,,\ r=1,2\,.$
From now on I will indicate with 
${\rm a}_{\,\imath}^{\,\dagger}\,,\,{\rm a}_{\,\imath}\ (\,\imath\,\in\mathfrak Q\,)$
the creation and destruction operators for particles of negative electric charge
$q=-\,e\ (e>0)$ while
${\rm b}_{\,\jmath}^{\,\dagger}\,,\,{\rm b}_{\,\jmath}\ (\,\jmath\,\in\mathfrak Q\,)$
the corresponding operators for the antiparticles of positive electric charge $e\,.$
\subsection{Pairs Operators Algebra}
Consider the pairs annihilation and pairs production dimensionless operators
\beq
\Pi(z)\,\equiv\,\sum_{\jmath\,\in\mathfrak Q}z_{\,\jmath}\,
{\rm a}_{\,\jmath}\,{\rm b}_{\,\jmath}\,=\,
\sum_{\jmath\,\in\mathfrak Q}z_{\,\jmath}\,\Pi_{\,\jmath}\\
\Pi^{\,\dagger}(\bar z)\,\equiv\,\sum_{\imath\,\in\mathfrak Q}\bar z_{\,\imath}\,
{\rm b}_{\,\imath}^{\,\dagger}\;\!{\rm a}_{\,\imath}^{\,\dagger}\,=\,
\sum_{\imath\,\in\mathfrak Q}\bar z_{\,\imath}\,\Pi^{\,\dagger}_{\,\imath}
\eeq
where we use the short notation
\[
\sum_{\imath\,\in\mathfrak Q}=\int\rmd{\bf p}\sum_r\cdots
\qquad\quad(\,{\bf p}\in{\mathbb R}^n\,,\quad n=1,2,3\,)
\]
whereas $z_{\,\imath}\equiv z({\bf p},r,\ldots)$ 
are complex valued {\em dimensionless} functions
and such that
\[
{\mathcal V}_n\,(2\pi)^{-\,n}\sum_{\imath\,\in\mathfrak Q}\ z_{\,\imath}\,\bar z_{\,\imath}\,=\,\nu_o
\qquad\quad(\,n=1,2,3\,)
\]
is a {\em pure number} that will be named the {\sl characteristic number} 
of the given pairs distribution function 
$z\,\!(\,{\bf p},r,\ldots\,)\,,$ while ${\mathcal V}_n$ is the volume 
of a very large cubic box in the $n$th dimensional euclidean space.
The creation and destruction operators 
for particles $(\,{\rm a}_{\,\imath}^{\,\dagger}\,,\,{\rm a}_{\,\jmath}\,)$ and 
antiparticles $(\,{\rm b}_{\,\imath}^{\,\dagger}\,,\,{\rm b}_{\,\jmath}\,)$
satisfy the usual canonical anticommutation relations
\[
\{\,{\rm a}_{\,\imath}^{\,\dagger}\,,\,{\rm a}_{\,\jmath}\,\}
=\{\,{\rm b}_{\,\imath}^{\,\dagger}\,,\,{\rm b}_{\,\jmath}\,\}\,=\,\delta_{\,\imath\jmath}
\qquad\quad(\,\imath,\jmath\,\in\mathfrak Q\,)
\]
all the remaining ones being equal to zero.
The Fock vacuum is defined as usual
\beq
{\rm a}_{\,\imath}\,|\,0\,\rangle = 0 = \langle\,0\,|\,{\rm a}_{\,\imath}^{\,\dagger}\qquad
{\rm b}_{\,\imath}\,|\,0\,\rangle = 0 = \langle\,0\,|\,{\rm b}_{\,\imath}^{\,\dagger}
\qquad
(\,\forall\,\imath\,\in\mathfrak Q\,)
\eeq
In a quite general manner, if we denote by $Q_a\ (\,a=1,2,\ldots,A\,)$
all the conserved charges of the system which are allowed by the
background field configuration, i.e.
\[
Q_a=\sum_{\jmath\,\in\mathfrak Q} q_a\,\Big({\rm a}^{\,\dagger}_{\,\jmath}\,{\rm a}_{\,\jmath}
-{\rm b}^{\,\dagger}_{\,\jmath}\,{\rm b}_{\,\jmath}\Big)
\qquad\quad(\,a=1,2,\ldots,A\,)
\]
where e.g. $q_a\ (\,a=1,2,\ldots,A\,)$ are the particle charges
while $-\,q_a$ the antiparticle charges,
then for any 1-pair state ${\rm b}_{\,\imath}^{\,\dagger}\;\!{\rm a}_{\,\imath}^{\,\dagger}\,|\,0\,\rangle$ 
of definite quantum numbers $\imath\,\in\mathfrak Q$ we evidently find
\beq
Q_a\,{\rm b}_{\,\imath}^{\,\dagger}\;\!{\rm a}_{\,\imath}^{\,\dagger}\,|\,0\,\rangle=0
\qquad\quad(\;\forall\,{\,\imath}\in{\mathfrak Q}\,,\quad\,a=1,2,\ldots,A\,)
\eeq
More generally, from the commutation relations
\beq
\left[\,Q_a\,,\,\Pi(z)\,\right]&=&\sum_{\jmath\,\in\mathfrak Q}q_a\sum_{\imath\,\in\mathfrak Q}z_{\,\imath}
\left[\,{\rm a}^{\,\dagger}_{\,\jmath}\,{\rm a}_{\,\jmath}
-{\rm b}^{\,\dagger}_{\,\jmath}\,{\rm b}_{\,\jmath}\,,\,
{\rm a}_{\,\imath}\;\!{\rm b}_{\,\imath}\,\right]\no
&=& (\,-\,q_a\,)\sum_{\imath\,\in\mathfrak Q}z_{\,\imath}
\Big({\rm a}_{\,\imath}\,{\rm b}_{\,\imath}
+ {\rm b}_{\,\imath}\,{\rm a}_{\,\imath}\Big)\ =\ 0
\eeq
which obviously also imply
$\left[\,Q_a\,,\,\Pi^{\,\dagger}(\bar z)\,\right]\,=\,0\ (\,a=1,2,\ldots,A\,)\,,$
it follows that if we set
\beq
\Pi^{\,\dagger}(\bar z)|\,0\,\rangle\
\equiv\ \left|\,\bar z\,\right\rangle
\eeq
then we find
\beq
Q_a\,\left|\,\bar z\,\right\rangle\,=\,0
\qquad\quad(\,a=1,2,\ldots,A\,)
\eeq
which means that the generic 1-pair state of distribution function $z\,\!(\,{\bf p},r,\ldots\,)$
is a common null eigenstate of all the conserved
charge operators. For example, in the case of Dirac spinors in the uniform electric field
the conserved charges are the electric charge, the three components of momentum
and the helicity, so that $A=5\,.$

The pairs creation and annihilation operators satisfy the commutation relations
\beq
\left[\,\Pi(z)\,,\,\Pi^{\,\dagger}(\bar z)\,\right]=
\sum_{\imath\,\in\mathfrak Q} z_{\,\imath}\sum_{\jmath\,\in\mathfrak Q}\bar z_{\,\jmath}\,
\left[\,{\rm a}_{\,\imath}\,{\rm b}_{\,\imath}\,,\,
{\rm b}^{\,\dagger}_{\,\jmath}\,{\rm a}^{\,\dagger}_{\,\jmath}\,\right]\no
=\;\sum_{\imath\,\in\mathfrak Q} z_{\,\imath}\sum_{\jmath\,\in\mathfrak Q}\bar z_{\,\jmath}\,\Big(
{\rm a}_{\,\imath}\left[\,{\rm b}_{\,\imath}\,,\,
{\rm b}^{\,\dagger}_{\,\jmath}\,{\rm a}^{\,\dagger}_{\,\jmath}\,\right]
+\left[\,{\rm a}_{\,\imath}\,,\,
{\rm b}^{\,\dagger}_{\,\jmath}\,{\rm a}^{\,\dagger}_{\,\jmath}\,\right]{\rm b}_{\,\imath}\Big)\no
=\;\sum_{\imath\,\in\mathfrak Q} z_{\,\imath}\sum_{\jmath\,\in\mathfrak Q}\bar z_{\,\jmath}\,\Big(
{\rm a}_{\,\imath}\,{\rm a}^{\,\dagger}_{\,\jmath} - 
{\rm b}^{\,\dagger}_{\,\jmath}\,{\rm b}_{\,\imath}\Big)\,\delta_{\,\imath\jmath}\no
=\;\sum_{\imath\,\in\mathfrak Q} z_{\,\imath}\,\bar z_{\,\imath}\Big({\rm a}_{\,\imath}\,{\rm a}^{\,\dagger}_{\,\imath}
- {\rm b}^{\,\dagger}_{\,\imath}\,{\rm b}_{\,\imath}\Big)\ \equiv\ -\,2{\rm N}(\bar zz)
\eeq
in which
\beq
{\rm N}(\bar zz) &\equiv&
\sum_{\imath\,\in\mathfrak Q} \textstyle\frac12 z_{\,\imath}\,\bar z_{\,\imath}
\Big({\rm b}^{\,\dagger}_{\,\imath}\,{\rm b}_{\,\imath}
-{\rm a}_{\,\imath}\,{\rm a}^{\,\dagger}_{\,\imath}\Big)\no
&=&\sum_{\imath\,\in\mathfrak Q} z_{\,\imath}\,\bar z_{\,\imath}\,n_{\,\imath}
\,=\, {\rm N}^{\;\!\dagger}(z\bar z)\no
&=&\sum_{\imath\,\in\mathfrak Q} \textstyle\frac12 z_{\,\imath}\,\bar z_{\,\imath}
\Big({\rm a}^{\,\dagger}_{\,\imath}\,{\rm a}_{\,\imath}
+ {\rm b}^{\,\dagger}_{\,\imath}\,{\rm b}_{\,\imath}\Big) - \frac12\,\nu_o
\eeq
Finally for $\nu_{\,\imath}\in{\mathbb R}\ (\,\forall\,\imath\,\in\mathfrak Q\,)$ we get
\beq
\left[\,{\rm N}(\nu)\,,\,\Pi(z)\,\right]&=&
\sum_{\jmath\,\in\mathfrak Q}{\textstyle\frac12}\,\nu_{\,\jmath}\sum_{\imath\,\in\mathfrak Q} z_{\,\imath}
\left[\,{\rm b}^{\,\dagger}_{\,\jmath}\,{\rm b}_{\,\jmath}
+{\rm a}^{\,\dagger}_{\,\jmath}\,{\rm a}_{\,\jmath}\,\,,\,
{\rm a}_{\,\imath}\,{\rm b}_{\,\imath}\,\right]\no
&=& - \sum_{\imath\,\in\mathfrak Q} \nu_{\,\imath}\,z_{\,\imath}\,{\rm a}_{\,\imath}\,{\rm b}_{\,\imath}\
=\ -\,\Pi(z\nu)\\
\left[\,{\rm N}(\nu)\,,\,\Pi^{\,\dagger}(\bar z)\,\right]&=&
\sum_{\imath\,\in\mathfrak Q} \nu_{\,\imath}\,\bar z_{\,\imath}\,{\rm b}^{\,\dagger}_{\,\imath}\,{\rm a}^{\,\dagger}_{\,\imath}
=\ \Pi^{\,\dagger}(\nu\bar z)
\eeq
It follows therefrom that the above three operators
do satisfy the well known commutation relations
\beq
&&\left[\,{\rm N}(\nu)\,,\,\Pi^{\,\dagger}(\bar z)\,\right]\ =\ \Pi^{\,\dagger}(\nu\bar z)\no
&&\left[\,\Pi(z)\,,\,{\rm N}(\nu)\,\right]\ =\ \Pi(\nu z)\label{su(2)}\\
&&\left[\,\Pi^{\,\dagger}(\bar z)\,,\,\Pi(z)\,\right]\ =\ 2{\rm N}(\bar zz)\nn
\eeq
in which
\beq
\Pi^{\,\dagger}(\bar z)=J_+(\bar z)=J_x(u)+\rmi\,J_y(v)\\
\Pi(z)=J_-(z)=J_x(u)-\rmi\,J_y(v)\\
{\rm N}(\nu)=J(uv)
\eeq
where the threesome of  operators
\beq
J_x(u)\equiv\textstyle\frac12\Big(\Pi(z)+\Pi^{\,\dagger}(\bar z)\Big)\no
J_y(v)\equiv\frac{1}{2i}\,\Big(\Pi^{\,\dagger}(\bar z)-\Pi(z)\Big)\no
J_z(uv)={\rm N}(\nu)
\label{angularmomenta}
\eeq
are a basis of hermitean generators obeying the well known SU(2) Lie algebra
\beq
\left[\,J_a(u)\,,\,J_b(v)\,\right]=i\hbar\,\varepsilon_{\,abc}\,J_c(uv)\qquad\quad
(\,a,b,c=1,2,3\,)
\eeq
The quantum state $|\,\bar z\,\rangle\,,$ that represents a generic 1-pair state with a 
momentum distribution function $\bar z_{\,\imath}\ (\,\imath\,\in\mathfrak Q\,)$ does satisfy
\beq
|\,\bar z\,\rangle = \Pi^{\,\dagger}(\bar z)\,|\,0\,\rangle
\qquad\quad
\langle\,z\,| = \langle\,0\,|\,\Pi(z)
\eeq
and has the norm
\beq
\langle\,z\,|\,\bar z\,\rangle=\langle\,0\,|\,[\,\Pi(z)\,,\,\Pi^{\,\dagger}(\bar z)\,]\,|\,0\,\rangle
= -\,2\langle\,0\,|\,{\rm N}(\bar zz)\,|\,0\,\rangle = \nu_o
\eeq
\subsection{The Schwinger Formula from the Algebraic Approach}
The general feature that characterizes the pairs production and annihilation
processes in the presence of external background uniform fields is the existence of
a nonsingular Bogolyubov similarity transformation $\mathcal S\,,$ 
the generator of which is
acting on the Fock space according to
\beq
\begin{array}{c}
{\rm A}_{\imath}={\mathcal S}^{\,-1}\,{\rm a}_{\,\imath}\,{\mathcal S}
\equiv\,c_{\,1\imath}\,{\rm a}_{\,\imath}-c_{\,2\,\imath}^\ast\,{\rm b}^{\,\dagger}_{\imath}
\label{bogolyubov_bis}\\
{\rm B}^{\,\dagger}_{\imath}={\mathcal S}^{\,-1}\,{\rm b}_{\imath}^{\,\dagger}\,{\mathcal S}
\equiv\,c_{\,1\imath}^{\,\ast}{\rm b}^{\,\dagger}_{\imath} + c_{\,2\,\imath}\,{\rm a}_{\,\imath}
\end{array}
\eeq
where
\[
|\,c_{\,1\imath}\,|^{\,2} + |\,c_{\,2\,\imath}\,|^{\,2} = 1\qquad\quad
(\,\forall\,\imath\,\in\mathfrak Q\,)
\]
in such a manner that the canonical anticommutation relations 
$$
\{\,{\rm A}_{\imath}\,,\,{\rm B}^{\,\dagger}_{\jmath}\,\}=0=
\{\,{\rm a}_{\,\imath}\,,\,{\rm b}^{\,\dagger}_{\jmath}\,\}\,,\qquad\quad{\rm et\ cetera}
$$
keep unchanged thanks to the similarity nature of the invertible transformation $\mathcal S\,.$
It follows that we come to the two Fock spaces ${\mathfrak F}_{\,\rm in}$ and 
${\mathfrak F}_{\,\rm out}$ which are generated by the cyclic vacuum states normalized to one and 
defined by 
\beq
{\rm a}_{\,\imath}\,|\,0\;{\rm in}\,\rangle={\rm b}_{\,\imath}\,|\,0\;{\rm in}\,\rangle=0\quad\qquad
(\,\forall\,\imath\,\in\mathfrak Q\,)\label{vacuum_in}\\
{\rm A}_{\,\jmath}\,|\,0\;{\rm out}\,\rangle={\rm B}_{\,\jmath}\,|\,0\;{\rm out}\,\rangle=0\quad\qquad
(\,\forall\,\jmath\,\in\mathfrak Q\,)
\label{vacuum_out}
\eeq
Now we have for example
\beq
{\rm A}_{\,\imath}\,{\rm a}^{\,\dagger}_{\,\imath}\,|\,0\;{\rm in}\,\rangle &=&
c_{\,1\imath}\,{\rm a}_{\,\imath}\,{\rm a}^{\,\dagger}_{\,\imath}\,|\,0\;{\rm in}\,\rangle
-c_{\,2\,\imath}^\ast\,{\rm b}^{\,\dagger}_{\imath}\,{\rm a}^{\,\dagger}_{\,\imath}\,|\,0\;{\rm in}\,\rangle\no
&=& {\mathcal V}_n\,(2\pi)^{-\,n}\,c_{\,1\imath}\,\,|\,0\;{\rm in}\,\rangle
-\,c_{\,2\,\imath}^\ast\,\Pi^{\,\dagger}_{\,\imath}\,|\,0\;{\rm in}\,\rangle
\eeq
so that
\beq
\langle\,{\rm in}\;0\,|\,{\rm A}_{\,\imath}\,{\rm a}^{\,\dagger}_{\,\imath}\,|\,0\;{\rm in}\,\rangle 
={\mathcal V}_n\,(2\pi)^{-\,n}\,c_{\,1\imath}\quad\qquad
(\,\forall\,\imath\,\in\mathfrak Q\,)
\label{vacuum_persistence}
\eeq
whence it follows that, as expected, the Bogolyubov coefficient $c_{\,1\imath}$
is nothing but the probability amplitude  that a pair of quantum numbers $\imath\,\in\mathfrak Q$
is {\bf not created} out of the vacuum or {\bf not absorbed} into the vacuum,
i.e. the relative vacuum persistence probability density.
In a similar way, by taking the in vacuum expectation value
\beq
\fl\qquad
(2\pi)^n\,{\mathcal V}_n^{\,-1}\,\langle\,{\rm in}\;0\,|\,\Pi_{\,\imath}\,
{\rm A}_{\,\imath}\,{\rm a}^{\,\dagger}_{\,\imath}\,|\,0\;{\rm in}\,\rangle
&=& -\,c_{\,2\,\imath}^\ast\,(2\pi)^n\,{\mathcal V}_n^{\,-1}\,\langle\,{\rm in}\;0\,|\,\Pi_{\,\imath}\,\Pi^{\,\dagger}_{\,\imath}\,|\,0\;{\rm in}\,\rangle\no
\fl\qquad
&=& -\,{\mathcal V}_n\,(2\pi)^{-\,n}\,c_{\,2\,\imath}^\ast\quad\qquad
(\,\forall\,\imath\,\in\mathfrak Q\,)
\eeq
it is also clear that we can understand the Bogolyubov coefficient $c_{\,2\imath}$
as the probability amplitude  that a pair of quantum numbers $\imath\,\in\mathfrak Q$
is created out of the vacuum or absorbed into the vacuum.
To proceed further on, let me define
\beq
{\mathcal S}(\,\theta,\widehat{\bf n})\;\equiv\;\exp\{-\,i\,\theta\cdot{\rm T}(z,\bar z,\nu)\}\\
\theta\cdot{\rm T}(z,\bar z,\nu)\equiv\,\Pi^{\,\dagger}(\,\theta\,\bar z)+\Pi(z\theta)+2{\rm N}(\theta\nu)
\eeq
where $\theta_{\ell}\ (\,\forall\,\ell\,\in\mathfrak Q\,)$ is a real functional parameter,
while the functional unit vector
$\widehat{\bf n}_{\,\imath}$ is related to the functional parameters 
$z_{\,\imath},\bar z_{\,\imath},\nu_{\,\imath}$ through the relationship
\[
\widehat{\bf n}_{\,\imath}^{\;\!2}=z_{\,\imath}\,\bar z_{\,\imath}+\nu_{\,\imath}^{\;\!2}=1\qquad\quad
(\,\forall\,\imath\,\in\mathfrak Q\,)
\]
For example, a suitable functional parametrization is provided by a pair of polar angles,
latitude $\Theta_{\,\imath}$ and azimuth $\phi_{\,\imath}\,,$ in such a manner to set
\[
\nu_{\,\imath}=\cos\Theta_{\,\imath}\qquad\quad 
z_{\,\imath}=\sin\Theta_{\,\imath}\,\exp\{-\,\rmi\phi_{\,\imath}\}\
\]
From the basic commutation relation
\beq
\fl
[\,{\rm T}(z,\bar z,\nu)\,,\,{\rm a}_{\,\imath}\,]=
-\,\bar z_{\,\imath}\,{\rm b}_{\imath}^{\,\dagger}-\nu_{\,\imath}\,{\rm a}_{\,\imath}\qquad
[\,{\rm T}(z,\bar z,\nu)\,,\,{\rm b}^{\,\dagger}_{\imath}\,]=-\,z_{\,\imath}\,{\rm a}_{\,\imath}
+ \nu_{\,\imath}\,{\rm b}^{\,\dagger}_{\imath}
\eeq
we readily obtain
\beq
[\,{\rm T}\,,\,[\,{\rm T}\,,\,{\rm a}_{\,\imath}\,]\,]={\rm a}_{\,\imath}\qquad
[\,{\rm T}\,,\,[\,{\rm T}\,,\,{\rm b}^{\,\dagger}_{\,\imath}\,]\,]={\rm b}^{\,\dagger}_{\,\imath}
\eeq
As a consequence, we actually obtain the most general Bogolyubov transformations in the form
\beq
\fl
{\rm A}_{\,\imath}={\rm a}_{\,\imath}+\rmi\,\theta_{\,\imath}\,[\,{\rm T}\,,\,{\rm a}_{\,\imath}\,]
+\textstyle{1\over2!}\,(\rmi\,\theta_{\,\imath})^2\,[\,{\rm T}\,,\,[\,{\rm T}\,,\,{\rm a}_{\,\imath}\,]\,]
+ {1\over3!}\,(\rmi\,\theta_{\,\imath})^3\,[\,{\rm T}\,,\,[\,{\rm T}\,,\,[\,{\rm T}\,,\,{\rm a}_{\,\imath}\,]\,]\,]\no
+\ \cdots\;=\;(\cos\theta_{\,\imath} -\rmi\,\nu_{\,\imath}\,\sin\theta_{\,\imath})\,{\rm a}_{\,\imath}\,-\,
\rmi\,{\rm b}_{\,\imath}^{\,\dagger}\,\bar z_{\,\imath}\sin\theta_{\,\imath}\no
=\;c_{\;\!1\,\imath}\,{\rm a}_{\,\imath}\;-\;c_{2\,\imath}^\ast\,{\rm b}^{\,\dagger}_{\,\imath}\\
\fl
{\rm B}^{\,\dagger}_{\,\jmath}={\rm b}^{\,\dagger}_{\,\jmath} +
\rmi\,\theta_{\jmath}\,[\,{\rm T}\,,\,{\rm b}^{\,\dagger}_{\,\jmath}\,]
+\textstyle{1\over2!}\,(\rmi\,\theta_{\jmath})^2\,[\,{\rm T}\,,\,[\,{\rm T}\,,\,{\rm b}^{\,\dagger}_{\,\jmath}\,]\,]
+ {1\over3!}\,(\rmi\,\theta_{\jmath})^3\,[\,{\rm T}\,,\,[\,{\rm T}\,,\,[\,{\rm T}\,,\,{\rm b}^{\,\dagger}_{\,\jmath}\,]\,]\,]\no
+\;\cdots\ =\ \left(\cos\theta_{\jmath}+\rmi\,\nu_{\,\jmath}\,
\sin\theta_{\jmath}\right)\,{\rm b}^{\,\dagger}_{\,\jmath}\,-\,\rmi\,{\rm a}_{\,\jmath}\,
z_{\,\jmath}\sin\theta_{\jmath}\no
=\;c_{1\,\jmath}^\ast\,{\rm b}^{\,\dagger}_{\,\jmath}\;+\;c_{2\,\jmath}\,{\rm a}_{\,\jmath}
\eeq
with
\beq
c_{\;\!1\,\jmath}\;\equiv\;\cos\theta_{\jmath} -\rmi\,\nu_{\,\jmath}\,\sin\theta_{\jmath}\qquad\quad
c_{\;\!2\,\jmath}^{\,\ast}\;\equiv\;\rmi\,\bar z_{\,\jmath}\sin\theta_{\jmath}
\eeq
\[
\widehat{\bf n}_\jmath^{\;\!2}\,=\,\bar z_\jmath\,z_\jmath +
\nu_\jmath^{\;\!2}\;=\;1\qquad\quad
|\,c_{\;\!1\,\jmath}\,|^{\;\!2}\;+\;|\,c_{\;\!2\,\jmath}\,|^{\;\!2}\ =\ 1
\]
It follows thereby that the functional unitary operator
\beq
{\mathcal S}(\theta,z,\nu)={\mathcal S}(\,\theta,\widehat{\bf n}\,)=
{\mathcal S}(\,c_1,c_2\,)\qquad\quad{\mathcal S}^{\,-1}={\mathcal S}^{\,\dagger}
\eeq
does generate the Bogolyubov similarity transformations which connect the extended in and out states and
fields according to the suitable definitions
\beq
\Psi_{\rm out}(x)\;=\;
{\mathcal S}^{\,-1}\,\Psi(x)\,
{\mathcal S}\qquad\quad
\Psi_{\rm in}(x)\;=\;
{\mathcal S}\,\Psi(x)\,
{\mathcal S}^{\,-1}\no
|\,{\rm out}\,\rangle\;=\;{\mathcal S}^{\,-1}\,|\,{\rm in}\,\rangle
\qquad\quad
|\,{\rm in}\,\rangle\;=\;{\mathcal S}\,|\,{\rm out}\,\rangle
\eeq
Moreover we obtain
\beq
\Psi_{\rm out}(x)&=&
\sum_{\imath\,\in\mathfrak Q}\;{\mathcal S}^{\,-1}
\left[\,{\rm a}_{\,\imath}\;u_{\,\imath\,-}(x) +\
{\rm b}^{\,\dagger}_{\,\imath}\;v_{\imath\,-}(x)\,\right]\,
{\mathcal S}\no
&=&\sum_{\jmath\,\in\mathfrak Q}\;
\left[\,{\rm A}_{\,\jmath}\;u_{\,\jmath\,-}(x) +\
{\rm B}^{\,\dagger}_{\,\jmath}\;v_{\jmath\,-}(x)\,\right]\\
\Psi_{\rm in}(x)&=&
\sum_{\imath\,\in\mathfrak Q}\;{\mathcal S}^{\,-1}
\left[\,{\rm A}_{\,\imath}\;u_{\,\imath\,+}(x) +\
{\rm B}^{\,\dagger}_{\,\imath}\;v_{\imath\,+}(x)\,\right]\,
{\mathcal S}\no
&=&\sum_{\jmath\,\in\mathfrak Q}\;
\left[\,{\rm a}_{\,\jmath}\;u_{\,\jmath\,+}(x) +\
{\rm b}^{\,\dagger}_{\,\jmath}\;v_{\jmath\,+}(x)\,\right]
\eeq
where, in the case of the spinor QED in a uniform electric field in four
spacetime dimensions we have e.g.
\beq
u_{\,\imath\,\pm}(x)\equiv u_{\,{\bf p}\,,\,r}^{(\pm)}\;\!(t,{\bf r})
\qquad\quad
v_{\,\jmath\,\pm}(x)\equiv v_{\,{\bf p}\,,\,r}^{(\pm)}\;\!(t,{\bf r})
\eeq
Hence, the most general Bogolyubov transformation is nothing but
a functional rotation in the Fock space with parameter functions
$(\theta,z,\nu)=(\theta,\widehat{\bf n})=
(c_1,c_2)$, the generators of which are the pairs emission $\Pi^{\,\dagger}(\theta\bar z)\,,$
the pairs absorption $\Pi(\theta z)$ and the pairs number ${\rm N}(\theta\nu)$ operators,
which actually fulfill the functional
commutation relations (\ref{su(2)}) arising from the SU(2) Lie algebra.

Suppose that at a very remote time ${\rm t}_{\,\imath}\;\to\;-\,\infty$ the system is in a definite state,
e.g. the vacuum $|\,0\;{\rm in}\,\rangle$ for instance, so that it contains no physical fermion particles.
The final state at a very future time ${\rm t}_{\,f}\;\to\;+\,\infty$
has some calculable probability to contain zero, one, two, etc. emitted pairs of fermion particles 
and antiparticles. For example the probability amplitude to remain in the vacuum state,
i.e., the probability amplitude of emitting no pairs, is given by
\beq
\fl\qquad
\langle\,0\;{\rm out}\,|\,0\;{\rm in}\,\rangle=
\langle\,0\;{\rm in}\,|\,{\mathcal S}\,|\,0\;{\rm in}\,\rangle=
\langle\,0\;{\rm out}\,|\,{\mathcal S}\,|\,0\;{\rm out}\,\rangle
\eeq
For this interpretation to make sense, one has to actually verify that the vacuum to vacuum probability
$$
W_{0\,,\,f\,\leftarrow\,\imath}\;\equiv\;|\,\langle\,0\;{\rm out}\,|\,0\;{\rm in}\,\rangle\,|^{\;\!2}
$$
is not greater than one.
To this concern consider the hermitean operator
\beq
N_{\imath}\;&\equiv&\;{\textstyle\frac12}\left(\,{\rm A}^{\,\dagger}_{\,\imath}\,{\rm A}_{\,\imath}-
{\rm B}_{\,\imath}\,{\rm B}^{\,\dagger}_{\,\imath}\,\right)\,=\,
{\mathcal S}^{\,-1}\,n_{\,\imath}\,{\mathcal S}\label{ngrande}\\
&=&{\textstyle\frac12}\left(\,c_{\;\!1\,\imath}^\ast\,{\rm a}^{\,\dagger}_{\,\imath}\; -\;
c_{\;\!2\,\imath}\,{\rm b}_{\,\imath}\,\right)
\left(\,c_{\;\!1\,\imath}\,{\rm a}_{\,\imath}\;-\;c_{\;\!2\,\imath}^\ast\,{\rm b}^{\,\dagger}_{\,\imath}\,\right)\no
&-&{\textstyle\frac12}\left(\,c_{\;\!1\,\imath}\,{\rm b}_{\,\imath}\;+\;
c_{\;\!2\,\imath}^\ast\,{\rm a}^{\,\dagger}_{\,\imath}\,\right)
\left(\,c_{\;\!1\,\imath}^\ast\,{\rm b}^{\,\dagger}_{\,\imath}\;+\;
c_{\;\!2\,\imath}\,{\rm a}_{\,\imath}\,\right)\no
&=& \left(\,|\,c_{\;\!1\,\imath}\,|^{\;\!2}\;-\;|\,c_{\;\!2\,\imath}\,|^{\;\!2}\,\right)n_{\,\imath}
+ c_{\;\!1\,\imath}^\ast\,c_{\;\!2\,\imath}^\ast\,\Pi^{\,\dagger}_{\,\imath}
+ c_{\;\!1\,\imath}\,c_{\;\!2\,\imath}\,\Pi_{\,\imath}\nn
\eeq
where 
\[
4\,|\,c_{\;\!1\,\jmath}\,c_{\;\!2\,\jmath}\,|^{\;\!2} +
\left(\,|\,c_{\;\!2\,\jmath}\,|^{\;\!2}-|\,c_{\;\!1\,\jmath}\,|^{\;\!2}\,\right)^2=1
\]
It follows that if we set
\beq
\zeta_{\,\jmath}\;\equiv\;2\,c_{\;\!1\,\jmath}\,c_{\;\!2\,\jmath}\;,\qquad\quad
w_\jmath\;\equiv\;|\,c_{\;\!2\,\jmath}\,|^{\;\!2} - |\,c_{\;\!1\,\jmath}\,|^{\;\!2}
\eeq
we can write the following {\sl golden operator identity} that is valid 
$\forall\,\imath,\jmath,\ell,\ldots\,\in\mathfrak Q\,:$

\beq
\fl
\qquad
2\,N_{\imath}\;=\;\zeta_{\,\imath}\,\Pi_{\,\imath} +
\bar\zeta_{\,\imath}\,\Pi^{\,\dagger}_{\,\imath} + 2w_{\,\imath}\,n_{\imath}
\qquad\quad
\bar\zeta_{\,\imath}\,\zeta_{\,\imath} + w_{\,\imath}^{\;\!2}=1
\label{golden}
\eeq
It is important to remark that the above equality (\ref{golden})
holds true thanks to the unitarity property satisfied by the 
Bogolyubov coefficients $c_{\;\!1\,\imath}\,,\,c_{\;\!2\,\jmath}\,.$
Then, {\bf for any complex distribution function} $\varphi\;\!({\bf p},r,\ldots)
=\varphi_{\,\imath}\,,$ we can write 
\beq
2\,{\mathcal N}(\varphi) &\equiv&
\sum_{\imath\,\in\mathfrak Q}\;2N_{\,\imath}\,\varphi_{\,\imath}\,=\,
\sum_{\imath\,\in\mathfrak Q}\;
\left(\,{\rm A}^{\,\dagger}_{\,\imath}\,{\rm A}_{\,\imath}-
{\rm B}_{\,\imath}\,{\rm B}^{\,\dagger}_{\,\imath}\,\right)\varphi_{\,\imath}\no 
&=&\Pi^{\,\dagger}(\bar\zeta\varphi)+\Pi(\zeta\varphi)+2J(\varphi w)\,=\,
\varphi\,\cdot\,{\rm T}(\zeta,\bar\zeta,w)
\eeq
where, for example, 
\[
w_{\,\imath}=\cos\Theta_{\,\imath}\qquad\quad
\zeta_{\,\imath}=\sin\Theta_{\,\imath}\,\exp\{-\,\rmi\,\phi_{\,\imath}\}
\qquad\quad(\,\forall\,\imath\,\in\,\mathfrak Q\,)
\]
Now, owing to
\beq
2\,N_{\imath}\,|\,0\;{\rm out}\,\rangle\,
&=&\{{\rm B}_{\,\imath}\,,\,{\rm B}^{\,\dagger}_{\,\imath}\}\,|\,0\;{\rm out}\,\rangle
=\delta_{\imath\imath}\,|\,0\;{\rm out}\,\rangle\no
&=&\delta^{\,(n)}({0})\,|\,0\;{\rm out}\,\rangle
\equiv\,{\mathcal V}_n\,(2\pi)^{-\,n}\,|\,0\;{\rm out}\,\rangle
\eeq
where ${\mathcal V}_n$ denotes the total volume occupied by a large box in the $n$th dimensional space
$(\,n=1,2,3\,)$, then
we eventually obtain
\beq
\langle\,0\;{\rm out}\,|\,0\;{\rm in}\,\rangle
&=&\langle\,0\;{\rm out}\,|\,{\mathcal S}_{\,\varphi}\,|\,0\;{\rm out}\,\rangle\no
&=&\langle\,0\;{\rm out}\,|\exp\left\{-\,2i\,{\mathcal N}(\varphi)\right\}|\,0\;{\rm out}\,\rangle\no
&=&\prod_{\,\imath\,\in\mathfrak Q}\;
\exp\left\lbrace-\,i\,\varphi_{\,\imath}\,{\mathcal V}_n\,(2\pi)^{-\,n}\right\rbrace\no
&=&\exp\left\lbrace-\,i\,{\mathcal V}_n\,(2\pi)^{-\,n}
\textstyle\sum_{\,\imath\,\in\mathfrak Q}\;\varphi_{\,\imath}\,\right\rbrace
\eeq
However, according to the natural interpretation
which arises from equation (\ref{vacuum_persistence}),
it is mandatory for consistency to identify
\beq
\varphi_{\,\jmath}&\equiv&
i\,\ln\,c_{\,1\jmath}^{\,\ast}
={\rm Arg}\,c_{\,1\jmath}+{\textstyle\frac12}\,i\ln\,|\,c_{\,1\jmath}\,|^{\;\!2}
\qquad\quad
(\,\forall\,\jmath\,\in\,{\mathfrak Q}\,)
\label{varphi_c1}
\eeq
As a matter of fact, according to Nikishov \cite{nikishov}, the logarithm of
vacuum to vacuum transition amplitude is provided by
\beq
\ln\,\langle\,0\;{\rm out}\,|\,0\;{\rm in}\,\rangle &=&
\sum_{\,\imath\,\in\mathfrak Q}\;\langle\,0_{\,\imath}\;{\rm out}\,|\,0_{\,\imath}\;{\rm in}\,\rangle\no
&=&\sum_{\,\imath\,\in\mathfrak Q}\;c_{\;\!1\,\imath}^{\,\ast}
\,=\,{\rm Tr}\ln c_{\;\!1}^{\,\ast}\no
&=&{\mathcal V}_n\,(2\pi)^{-\,n}
\sum_{\,\imath\,\in\mathfrak Q}\;\ln c_{\;\!1\,\imath}^{\,\ast}
\eeq 
A close comparison evidently yields once again (\ref{varphi_c1}).
As a consequence it is possible to express the out vacuum in terms of the in operators
in the explicit form
\beq
|\,0\;{\rm out}\,\rangle &=& {\mathcal S}_{\,\varphi}^{\,-1}\,|\,0\;{\rm in}\,\rangle\no
&=&
\exp\left\{2i\,{\mathcal N}(\varphi)\right\}\,|\,0\;{\rm in}\,\rangle\no
&=& \exp\left\{-\,2\,{\mathcal N}(\,\ln\,c_{\,1}^{\,\ast}\,)\right\}\,|\,0\;{\rm in}\,\rangle\no
&=&\exp\left\{\textstyle\sum_{\jmath\,\in\mathfrak Q}\;
{\rm B}_{\,\jmath}\,{\rm B}_{\,\jmath}^{\,\dagger}\,\ln\,c_{\,1\jmath}^{\,\ast}\right\}\no
&\times&\exp\left\{-\textstyle\sum_{\imath\,\in\mathfrak Q}\;
{\rm A}^{\,\dagger}_{\,\imath}\,{\rm A}_{\,\imath}\,\ln\,c_{\,1\imath}^{\,\ast}\right\}\,
|\,0\;{\rm in}\,\rangle
\eeq
Then from the commutation relations 
\beq
[\,{\rm A}^{\,\dagger}_{\,\imath}\,{\rm A}_{\,\imath}\,,\,{\rm A}_{\,\varkappa}\,]\,
=\,-\,{\rm A}_{\,\imath}\,\delta_{\,\imath\varkappa}\qquad\quad
[\,{\rm B}_{\,\imath}\,{\rm B}^{\,\dagger}_{\,\imath}\,,\,{\rm A}_{\,\varkappa}\,]\,
=\,0
\eeq
it immediately follows that, by the very construction,
\beq
{\rm A}_{\,\varkappa}\,|\,0\;{\rm out}\,\rangle\,=\,0\qquad\quad
(\,\forall\,\varkappa\,\in\,{\mathfrak Q}\,)
\eeq
and in a quite analogous way one can readily check the other relation
\beq
{\rm B}_{\,\jmath}\,|\,0\;{\rm out}\,\rangle\,=\,0\qquad\quad
(\,\forall\,\jmath\,\in\,{\mathfrak Q}\,)
\eeq
and the corresponding ones under the exchange of in and out.
Notice that the invertible operator ${\mathcal S}_{\,\varphi}$ 
is not unitary, owing to the presence of an imaginary part in
the distribution function $\varphi=i\ln\,c_{\,1}^{\,\ast}\,.$
Furthermore, from the golden operator identity (\ref{golden})
it follows that we can write
\beq
2\,{\mathcal N}(\varphi)=
\varphi\cdot{\rm T}(\zeta,\bar\zeta,w)\equiv\,
\Pi^{\,\dagger}(\,\varphi\,\bar\zeta)+\Pi(\zeta\varphi)+2{\rm N}(\varphi w)
\eeq
in such a manner that the out vacuum state can be expressed {\em \`a la} Dirac
as an infinite sea of pairs, i.e., a coherent like state involving any number
of pairs of any quantum numbers: namely,
\beq
&&|\,0\;{\rm out}\,\rangle={\mathcal S}_{\,\varphi}^{\,-1}\,|\,0\;{\rm in}\,\rangle=
\exp\left\{-\,2\;{\mathcal N}(\,\ln\,c_{\,1}^{\,\ast}\,)\right\}\,|\,0\;{\rm in}\,\rangle\no
&&= \exp\left\{-\,\Pi^{\,\dagger}(\,\bar\zeta\ln\,c_{\,1}^{\,\ast}\,)-\,
\Pi(\,\zeta\,\ln\,c_{\,1}^{\,\ast}\,)-\,
2{\rm N}(\,w\,\ln\,c_{\,1}^{\,\ast}\,)
\right\}\,|\,0\;{\rm in}\,\rangle
\eeq
in which
\[
\fl\qquad\quad
\zeta_{\,\jmath}\;\equiv\;2\,c_{\;\!1\,\jmath}\,c_{\;\!2\,\jmath}\qquad\quad
w_\jmath\;\equiv\;|\,c_{\;\!2\,\jmath}\,|^{\;\!2} - |\,c_{\;\!1\,\jmath}\,|^{\;\!2}
\qquad\quad(\,\forall\,\jmath\in{\mathfrak Q}\,)
\]

Finally, as an example, in the case of spinor QED in the presence of a uniform electric field
on the four dimensional Minkowski spacetime, i.e. $n=3\,,$ we have
\beq
\fl\quad
\langle\,0\;{\rm out}\,|\,0\;{\rm in}\,\rangle=
\exp\left\lbrace -\,i\,{\mathcal V}\,({\rm t}_f-{\rm t}_{\;\!\imath})\;
2\cdot{eE\over8\pi^3}\int\rmd^2p_\perp\;\varphi\;\!(\lambda)\right\rbrace
\eeq
with
\[
{\mathcal V}_{\,3}={\mathcal V}
\qquad\quad
p_\perp\,\equiv\,(\,p_y,p_z\,)
\qquad\quad
\lambda\,\equiv\,(\,p_\perp^{\,2}+M^2\,)/eE
\]
in such a manner that we can write 
\beq
\fl\qquad
|\,\langle\,0\;{\rm in}\,|\,0\;{\rm out}\,\rangle\,|^{\;\!2}\;
&=& \exp\left\lbrace{\mathcal V}\,({\rm t}_f-{\rm t}_{\;\!\imath})\,
{eE\over(2\pi)^3}\int\rmd^2 p_\perp\;\ln\,|\,c_{\,1}\;\!(\lambda)\,|^{\;\!2}
\right\rbrace\no
\fl\qquad
&=&\exp\left\lbrace{\mathcal V}\,({\rm t}_f-{\rm t}_{\;\!\imath})\,
{eE\over8\pi^3}\int\rmd^2 p_\perp\;\ln\left(1-\e^{-\,\pi\lambda}\right)\right\rbrace
\eeq
which is nothing but the Schwinger formula.
\section{Discussion and Conclusions}
In this paper I have shown that the processes of emission and absorption
of charged fermion-antifermion pairs, in the presence of a background uniform electric field,
can be fully described by means of an algebraic approach based upon the functional
SU(2) Lie algebra, as obeyed by the pairs operators. Actually, the threesome
of functional generators of SU(2) are nothing but the pair creation,
the pair destruction and the pairs number operators respectively.
This result allows to put the Schwinger
pair production mechanism within a rigorous framework of quantum field theory
in the presence of external, classical, background fields. In particular, the Bogolyubov
transformations leading to the Bogolyubov coefficients, nicely appears to be nothing but a 
functional rotation in the Fock space. The present derivation is strongly tailored to
the spinor QED, but is not difficult to imagine that the algebraic approach can
be generalized to other contexts, such as varying external fields \cite{dumlu}, charged scalar or vector 
quantized fields \textit{et cetera}. Last but not least, even the Bogolyubov trasformations
relating inertial and noninertial observers and leading to the famous Unruh and Hawking
effects could be revisited in the light of the algebraic approach. This might simplify
and clarify the main formul\ae, once the suitable functional Lie groups have been identified.
\section*{Acknowledgements}
I wish to acknowledge the support of the Istituto Nazionale di Fisica Nucleare,
Iniziativa Specifica PI13, that contributed to the successful completion of this project.
\section*{Appendix A: the Parabolic Cylinder Functions}

The parabolic cylinder functions, of the special form we are interested in the present context,
can be defined, e.g., by the integral representation
{\bf 9.241} 1. p. 1092 of ref.~\cite{gradshteyn}
\beq
D_{-i\lambda/2}\,[\,\pm(1+i)\,\xi\,] &=&
{1\over \surd\pi}\,2^{-i\lambda/2+1/2}\,\e^{-\pi\lambda/4}\,\e^{i\xi^2/2}
\nonumber\\
&\times& \int_{-\infty}^\infty x^{-i\lambda/2}\,\e^{-2x^2\pm 2ix(1+i)\xi}\ dx\,,
\label{parcyldef}
\eeq
where $\lambda>0\,,\ \xi\in{\mathbb R}\,,\ {\rm arg}\,x^{-i\lambda/2}=\lambda/2$ for $x<0\,,$ so that
\beq
D^{\,*}_{-i\lambda/2}\,[\,\pm(1+i)\,\xi\,] &=&
{1\over \surd\pi}\,2^{\,i\lambda/2+1/2}\,\e^{-\pi\lambda/4}\,\e^{-i\xi^2/2}
\nonumber\\
&\times& \int_{-\infty}^\infty x^{i\lambda/2}\,\e^{-2x^2\mp 2ix(1-i)\xi}\ dx\,.
\eeq
After the change of variable $x\,\longmapsto\,-\,x$
\beq
D^{\,*}_{-i\lambda/2}\,[\,\pm(1+i)\,\xi\,] &=&
{1\over \surd\pi}\,2^{\,i\lambda/2+1/2}\,\e^{\pi\lambda/4}\,\e^{-i\xi^2/2}
\nonumber\\
&\times& \int_{-\infty}^\infty x^{i\lambda/2}\,\e^{-2x^2\pm 2ix(1-i)\xi}\ dx\,,
\eeq
we eventually come to the conjugation property
\beq
D^{\,*}_{-i\lambda/2}\,[\,\pm (1+i)\,\xi\,]\ =\
D_{\,i\lambda/2}\,[\,\pm (1-i)\,\xi\,]\,,
\label{parcylconj}
\eeq
as na\"\i vely expected.
The following special values appear in our calculations,
\beq
&& D_{\,\pm i\lambda/2}\,(0)=\pi^{-1/2}\,2^{\pm i\lambda/4}\,
\Gamma\left(\frac12\pm{i\lambda\over 4}\right)\,\cosh{\pi\lambda\over 4}\,,\\
&& D_{\,\pm i\lambda/2-1}\,(0)=\pm i\,\pi^{-1/2}\,2^{\pm i\lambda/4-1/2}\,
\Gamma\left(\pm{i\lambda\over 4}\right)\,\sinh{\pi\lambda\over 4}\,;\\
&& \pm\,{\lambda\over 2}\,|\,D_{\,\pm i\lambda/2-1}\,(0)\,|^2 =
\pm\,\sinh{\pi\lambda\over 4}\,,\\
&& |\,D_{\,\pm i\lambda/2}\,(0)\,|^2=\cosh{\pi\lambda\over 4}\,.
\eeq
The parabolic cylinder functions fulfill the recursion formulas
\beq
&& \frac{d}{dz}\,D_\nu(z) = -\,\frac12\,zD_\nu(z)+\nu\,D_{\nu-1}(z)\,,
\label{recursionlowering}\\
&& \frac{d}{dz}\,D_\nu(z) = \frac12\,zD_\nu(z) - D_{\nu+1}(z)
\label{recursionraising}\,.
\eeq
Consider the combination
\beq
D_+ &\equiv&
D_{-i\lambda/2}\,[\,(1+i)\,\xi\,]\,D_{\,i\lambda/2}\,[\,(1-i)\,\xi\,]
\nonumber\\
&+& {\lambda\over 2}\,D_{-i\lambda/2-1}\,[\,(1+i)\,\xi\,]\,
D_{\,i\lambda/2-1}\,[\,(1-i)\,\xi\,]\,.
\label{D_+}
\eeq
From the recursion formul\ae\  we get
\beq
&& 2\,{d\over d\xi}\,D_{-i\lambda/2}\,[\,(1+i)\,\xi\,]\,
   D_{\,i\lambda/2}\,[\,(1-i)\,\xi\,]\ =\nonumber\\
&& \lambda\,(1+i)\,D_{-i\lambda/2}\,[\,(1+i)\,\xi\,]\,
   D_{\,i\lambda/2-1}\,[\,(1-i)\,\xi\,]\ +\ {\rm c.c.}\nonumber\\
&& \lambda\,{d\over d\xi}\,D_{-i\lambda/2-1}\,[\,(1+i)\,\xi\,]\,
   D_{\,i\lambda/2-1}\,[\,(1-i)\,\xi\,]\ =\nonumber\\
&& -\,\lambda\,(1+i)\,D_{-i\lambda/2}\,[\,(1+i)\,\xi\,]\,
   D_{\,i\lambda/2-1}\,[\,(1-i)\,\xi\,]\ +\ {\rm c.c.}\,,\nonumber
\eeq
so that the above combination $D_+$ does not depend upon $\xi$
and from the conjugation property (\ref{parcylconj}) we can write
\beq
D_+ =\ |D_{-i\lambda/2}(0)|^{\,2} +
{\lambda\over 2}\,|D_{-i\lambda/2-1}(0)|^{\,2}
=\ \exp\{\pi\lambda/4\}\,.
\label{parcylnorma}
\eeq
Let us now consider the further combination
\beq
D_- &\equiv&
D_{-i\lambda/2}\,[\,(1+i)\,\xi\,]\,D_{\,i\lambda/2}\,[\,-(1-i)\,\xi\,]
\nonumber\\
&-& {\lambda\over 2}\,D_{-i\lambda/2-1}\,[\,(1+i)\,\xi\,]\,
D_{\,i\lambda/2-1}\,[\,-(1-i)\,\xi\,]\,.
\label{D_-}
\eeq
From the recursion formul\ae\  we get
\beq
&& 2\,{d\over d\xi}\,\{D_{\,i\lambda/2}\,[\,(i-1)\,\xi\,]\,
D_{\,-i\lambda/2}\,[\,(i+1)\,\xi\,]\}\ =\nonumber\\
&& \lambda\,(1-i)\,D_{\,i\lambda/2}\,[\,(i-1)\,\xi\,]\,
   D_{\,-i\lambda/2-1}\,[\,(i+1)\,\xi\,]\ -\nonumber\\
&& -\,\lambda\,(1+i)\,D_{\,-i\lambda/2}\,[\,(i+1)\,\xi\,]\,
   D_{\,i\lambda/2-1}\,[\,(i-1)\,\xi\,]\;\nonumber\\
&& \lambda\,{d\over d\xi}\,\{D_{-i\lambda/2-1}\,[\,(i+1)\,\xi\,]\,
D_{\,i\lambda/2-1}\,[\,(i-1)\,\xi\,]\}\ =\nonumber\\
&& \lambda\,(1-i)\,D_{\,i\lambda/2}\,[\,(i-1)\,\xi\,]\,
   D_{\,-i\lambda/2-1}\,[\,(i+1)\,\xi\,]\ -\nonumber\\
&& -\,\lambda\,(1+i)\,D_{\,-i\lambda/2}\,[\,(i+1)\,\xi\,]\,
   D_{\,i\lambda/2-1}\,[\,(i-1)\,\xi\,]\,,
\eeq
which leads to the conclusion that also the quantity $D_-$
is independent of $\xi$, and yields
\beq
D_- =\ |D_{-i\lambda/2}(0)|^{\,2} -
{\lambda\over 2}\,|D_{-i\lambda/2-1}(0)|^{\,2}
=\ \exp\{-\pi\lambda/4\}\,.
\label{parcylpair}
\eeq
The above important properties of the parabolic cylinder functions
can be summarized in the remarkable formula
\beq
D_\pm =\ |D_{\,i\lambda/2}\,(0)|^{\,2} \pm
{\lambda\over 2}\,|D_{\,i\lambda/2-1}(0)|^{\,2}
=\ \exp\{\pm\pi\lambda/4\}\,.
\label{parcylfunda}
\eeq
Consider the second order differential equations
\beq
\left({d^2\over d\xi^{2}}+\xi^{2}+\lambda\pm\,i\right)f_\pm(\xi,\lambda)=0\,.
\eeq
Two pairs of linearly independent solutions for the upper sign equation are
\beq
f_+^{\,(1)}(\pm\,\xi,\lambda) = D_{-i\lambda/2}\,\left(\pm\,\xi\sqrt2\,\e^{\,\pi i/4}\right)\\
f_+^{\,(2)}(\pm\,\xi,\lambda) = D_{\,i\lambda/2-1}\,\left(\pm\,\xi\sqrt2\,\e^{\,-\,\pi i/4}\right)
\eeq
while two couples of linearly independent solutions for the lower sign equation are
\beq
f_-^{\,(1)}(\pm\,\xi,\lambda)=D_{-i\lambda/2-1}\,\left(\pm\,\xi\sqrt2\,\e^{\,\pi i/4}\right)\\
f_-^{\,(2)}(\pm\,\xi,\lambda)=D_{\,i\lambda/2}\,\left(\pm\,\xi\sqrt2\,\e^{\,-\,\pi i/4}\right)
\eeq
To the aim of verifying linear independence we have to compute the wronskian.
Let us first calculate derivatives by means of the recursion formul\ae\
(\ref{recursionlowering}) and (\ref{recursionraising}) that yield
\beq
\frac{d}{d\xi}\,D_{-i\lambda/2}\,\left(\pm\,\xi\sqrt2\,\e^{\,\pi i/4}\right) =\no
-\,i\,\xi\,D_{-i\lambda/2}\,\left(\pm\,\xi\sqrt2\,\e^{\,\pi i/4}\right)
\mp\,{\lambda\over\surd\,2}\,\e^{\,3\pi i/4}\,D_{-i\lambda/2-1}\,\left(\pm\,\xi\sqrt2\,\e^{\,\pi i/4}\right)\nn
\eeq
and thereby
\beq
W\left[\,f_+^{\,(1)}(\xi,\lambda)\,,\,f_+^{\,(1)}(-\,\xi,\lambda)\,\right]
&=& {1+i\over\surd\,\pi}\,\Gamma\left(-\,{i\lambda\over2}\right)\,\sinh\left({\pi\lambda\over2}\right)
\eeq
On the other side we readily find
\beq
W\left[\,f_+^{\,(1)}(\pm\,\xi,\lambda)\,,\,f_+^{\,(2)}(\pm\,\xi,\lambda)\,\right]\ =\
\mp\,(1-i)\,\exp\{\,\pi\lambda/4\}
\eeq
and analogous relationships for the other solutions.

\bigskip
In order to understand the physical meaning of the solutions
of the wave field equations,
we have to analyze the leading asymptotic behavior of
the parabolic cylinder functions.
Then
from eq.~{\bf 9.246} 1. p. 1093 of
ref.~\cite{gradshteyn} we have
\beq
\fl
\left.\begin{array}{c}
D_{-i\lambda/2}\,\left(\xi\sqrt2\,\e^{\,\pi i/4}\right)\ \sim\
(2\xi^2)^{\,-i\lambda/4}\,\e^{\pi\lambda/8}\exp\left\{-\,i\,\xi^2/2\right\}\\
D_{-i\lambda/2-1}\,\left(\xi\sqrt2\,\e^{\pi i/4}\right)\ \sim\
O(\xi^{-1})
\end{array}\right\rbrace\qquad\quad(\,\xi\gg\lambda>0\,)\nn
\label{+asyt-infty}
\eeq
If instead $\xi\ll-\lambda$ we have either
$\xi\,\e^{\pi i/4}=|\,\xi\,|\,\e^{5\pi i/4}$ or else
$\xi\,\e^{\pi i/4}=|\,\xi\,|\,\e^{-3\pi i/4}\,.$
Now, for ${\rm arg}(\xi\,\e^{\pi i/4})=5\pi i/4\,,$ no reliable
asymptotic expansion is available, so that from eq.~{\bf 9.246} 3.,
p. 1094 of ref.~\cite{gradshteyn} we obtain the {\it bona fide}
leading behaviour for $\xi\ll\,-\,\lambda<0$: namely,
\beq
&& D_{\,-i\lambda/2-1}\,\left(\xi\sqrt2\,\e^{\pi i/4}\right)\ =\
D_{\,-i\lambda/2-1}\,\left(|\,\xi\,|\sqrt2\,\e^{-3\pi i/4}\right)\nonumber\\
&& \sim\ \frac{\sqrt{2\pi}}{\Gamma(1+i\lambda/2)}\,(2\xi^2)^{\,i\lambda/4}\,
\exp\left\{-\,\frac{\pi\lambda}{8}+{i\xi^2\over 2}\right\}\,,
\label{+asyt+infty}\\
&& D_{\,-i\lambda/2}\,\left(|\,\xi\,|\sqrt2\,\e^{-3\pi i/4}\right)\ \sim\
(2\xi^2)^{\,-i\lambda/4}\,
\exp\left\{-\,\frac{3\pi\lambda}{8} - {i\xi^2\over 2}\right\}\nonumber\,.
\eeq
Of course, the situation becomes exactly time-reversed for the two other
linearly independent solutions: namely, for $\xi\gg\lambda>0$ we find
\beq
&& D_{-i\lambda/2-1}\,\left(-\xi\sqrt2\,\e^{\pi i/4}\right)\ =\
D_{\,-i\lambda/2-1}\,\left(\xi\sqrt2\,\e^{-3\pi i/4}\right)\nonumber\\
&& \sim\ \frac{\sqrt{2\pi}}{\Gamma(1+i\lambda/2)}\,(2\xi^2)^{\,i\lambda/4}\,
\exp\left\{-\,\frac{\pi\lambda}{8}+{i\xi^2\over 2}\right\}\,,\nonumber\\
&& D_{-i\lambda/2}\,\left(-\xi\sqrt2\,\e^{\pi i/4}\right)\ =\
D_{\,-i\lambda/2}\,\left(\xi\sqrt2\,\e^{-3\pi i/4}\right)\nonumber\\
&& \sim\ (2\xi^2)^{\,-i\lambda/4}\,
\exp\left\{-\,\frac{3\pi\lambda}{8} - {i\xi^2\over 2}\right\}\,,
\label{-asyt+infty}
\eeq
whereas for $\xi\ll -\lambda<0$ we obtain
\beq
\fl
\left.\begin{array}{c}
D_{-i\lambda/2}\,\left(|\xi|\sqrt2\,\e^{\,\pi i/4}\right)\ \sim\
(2\xi^2)^{\,-i\lambda/4}\,\e^{\pi\lambda/8}\exp\left\{-\,{i\,\xi^2/2}\right\}\\
D_{-i\lambda/2-1}\,\left(|\xi|\sqrt2\,\e^{\pi i/4}\right)\ \sim\
O(\xi^{-1})
\end{array}\right\rbrace\qquad\quad(\,\xi\ll-\,\lambda<0\,)\nn
\label{-asyt-infty}
\eeq
For a given particle momentum $p_x=p$
we shall associate the stationary asymptotic phase
\beq
\fl\qquad
{\textstyle\frac12}\,{\xi^2}(t)={\textstyle\frac12}\,eEt^2 - pt + {p^2/2eE}
\qquad\quad
{\textstyle\frac12}\,{\dot\xi^2}(t)=eEt-p\,\sim\,\omega t
\eeq
to the {\sl positive frequency solutions} $\omega t$
which describe either a particle, {i.e.} an electron of momentum $p$
and charge $-\,e$ when $t\to+\infty\,,$
or an antiparticle, {i.e.} a positron of momentum $-\,p$
and charge $e$ when $t\to-\infty\,.$
Conversely,
the {\sl negative frequency solutions} $-\,{\xi^2}(t)$
associated to the stationary asymptotic phases
\beq
\fl\
-\,{\textstyle\frac12}\,{\xi^2}(t)= -\,{\textstyle\frac12}\,eE\,ct^2 + pct - {p^2/2eE}
\qquad\quad
-\,{\textstyle\frac12}\,{\dot\xi^2}(t)= -\,eEt+p\,\sim\,-\,\omega t
\eeq
will instead describe either a particle of momentum $p$
and charge $-\,e$ or an antiparticle of momentum $-\,p$
and charge $e$ when $t\to\,-\,\infty\,.$
\section*{Appendix B : the Euclidean Formulation}
In order to make contact with the celebrated Schwinger's formula
(\ref{instanton}) it is expedient to turn to the euclidean formulation
\cite{coleman}.
To this aim, consider the euclidean Dirac operator for an electron
\beq
\EDirac + M \equiv 
\left(\partial_\mu - ie\,A_{\,\mu}\right)\gamma_\mu + M
\eeq
where
\beq
&& x_\mu=(\tau,x,y,z)=(it,x,y,z)\qquad A_{\,\mu}=(0,{\mathcal E}\tau,0,0)\\
&& \{\gamma_\mu,\gamma_\nu\} = 2\delta_{\mu\nu}\qquad
\gamma_\mu=\gamma_\mu^{\,\dagger}\qquad\gamma_\mu=(\,\gamma^0,-\,i\gamma^{\,k}\,)
\eeq
in which $\mathcal E\,=\,-\,iE$ is the euclidean electric field.
The euclidean Dirac operator is a normal elliptic operator so that one can
can safely define its complex power and the corresponding euclidean effective action in the 1--loop
approximation by means of the zeta function regularisation
\cite{zeta},\cite{blauSS}: namely,
\beq
{\cal S}^{\rm E}_{\rm eff}[A_{\,\mu}]&=&
{\cal S}^{\rm E}_{\rm cl}[A_{\,\mu}]-{\textstyle\frac12}\,\ln
{\rm det}\,[\,(\EDirac\,\EDirac^{\,\dagger}+M^2)/\mu^2\,]\nn\\
&=& \left.{\textstyle\frac12}\,{\mathcal E}^2 - {\rmd\over 2\rmd s}\,\right\rfloor_{s=0}
{\rm Tr}\,\left[\,(\EDirac\,\EDirac^{\,\dagger}+M^2)/\mu^2\,\right]^{-s}
\nn\\
{\cal S}^{\rm E}_{\rm cl}[A_\mu]&=& 
\int \rmd^4x\;{\overline\psi}(x)\,
\{\,\gamma_\rho\,[\,\partial_\rho - eA_\rho(x)\,]+M\,\}\,\psi(x)
\eeq
in which $\mu$ is a suitable reference mass scale. The euclidean second order 
differential operator turns out to be elliptic and reads
\beq
\EDirac\,\EDirac^{\,\dagger}+M^2=
-\,\partial_\tau^2+(\,p_x+e{\mathcal E}\tau)^2+p_y^2+p_z^2+M^2+e{\mathcal E}\alpha^1
\eeq
where, of course,
\beq
{\bf p} = -\,i\,\nabla\qquad\quad
\alpha^1=\left\lgroup\begin{array}{cc}
0 & \sigma_x \\
\sigma_x & 0\\
\end{array}\right\rgroup
\eeq 
Furthermore we have
\beq
[\,-\,\partial_\tau^2+(\,p_x+e{\mathcal E}\tau)^2+p_y^2+p_z^2+M^2,e{\mathcal E}\alpha^1\,]=0
\eeq
and we can easily find the spectrum and degeneracy 
of the second order differential scalar
operator $-\,\partial_\tau^2+(\,p_x+e{\mathcal E}\tau)^2+p_y^2+p_z^2+M^2$: namely,
\beq
&& \lambda_{\,n,\,p_y,\,p_z}=p_y^2+p_z^2+M^2+e{\mathcal E}(2n+1)
\nonumber\\
&& p_y,\,p_z \in {\mathbb R}\qquad 
n+1 \in {\mathbb N}\qquad \Delta={e{\mathcal E}/2\pi}
\eeq
It follows therefrom that we can write \cite{gradshteyn}
%
\beq
&& \left.{d\over ds}\,\right\rfloor_{s=0}
{\rm Tr}\,\left[\,(\EDirac\,\EDirac^{\,\dagger}+M^2)/\mu^2\,\right]^{-s}\
=\ {e^2{\mathcal E}^2\over4\pi^2}\,(\,{\rm vol}\,)\nonumber\\
&& \times\ \left.{d\over ds}\,\right\rfloor_{s=0}
\left({\mu^2\over eE}\right)^s\,{1\over\Gamma(s)}\int_0^\infty
dy\ y^{\,s-2} e^{-ay}\,\coth y
\nonumber\\
&& =\ (\,{\rm vol}\,)\;(\,e{\mathcal E}/\pi\,)^2
\nonumber\\
&& \times\ \left.{d\over ds}\,\right\rfloor_{s=0}
\left({\mu^2\over 2e{\mathcal E}}\right)^s\,
\frac{1}{s-1}\,
\left[\,\zeta\left(s-1,{a\over2}\right)
-\frac{a^2}{4}\left({a\over2}\right)^{-s}\,\right]
\nonumber
\eeq
with $a=M^2/e{\mathcal E}\,,\ \Re{\rm e}\,s>2\,,$ 
where we can identify (\,vol\,) $={\mathcal V\,\mathcal T}$ 
in which $\mathcal T=i\,(\mathrm t_f-\mathrm t_{\,i})$ is the total euclidean time,
so that,
taking the simplest renormalization prescription $\mu=M$ 
into account, we end up with
\beq
{\cal L}^{\rm E}_{\rm eff}&=&
{\textstyle\frac12}\,{\mathcal E}^2 - \left(\frac{M}{2\pi}\right)^2\ +\\ 
&+& {e^2{\mathcal E}^2\over2\pi^2}\left\lbrace
\left(1+\ln\frac{M^2}{2e{\mathcal E}}\right)
\zeta\left(-1,{M^2\over2e{\mathcal E}}\right)
- \zeta^{\,\prime}\left(-1,{M^2\over2e{\mathcal E}}\right)
\right\rbrace
\nonumber
\eeq

On the other hand, from the representation as a series of 
the Riemann zeta function
$\zeta(z,q)$ for $\Re{\rm e}\,z<0\,,\ 0<q\le 1\,,$ we have
\beq
\zeta(z,q)&=&\frac{2\,\Gamma(1-z)}{(2\pi)^{1-z}}
\sum_{n=1}^\infty
n^{z-1}\sin\left(2\pi nq+{z\pi\over2}\right)
\nonumber\\
\zeta^{\,\prime}(z,q)&=&\frac{2\,\Gamma(1-z)}{(2\pi)^{1-z}}\,
[\,\ln2\pi-\psi(1-z)\,]\sum_{n=1}^\infty
n^{z-1}\sin\left(2\pi nq+{z\pi\over2}\right)
\nonumber\\
&+&\frac{2\,\Gamma(1-z)}{(2\pi)^{1-z}}
\sum_{n=1}^\infty
n^{z-1}\,(\ln n)\sin\left(2\pi nq+{z\pi\over2}\right)
\nonumber\\
&+&\frac{2\,\Gamma(1-z)}{(2\pi)^{1-z}}\,
\sum_{n=1}^\infty
n^{z-1}\,{\pi\over2}\,\cos\left(2\pi nq+{z\pi\over2}\right)
\nonumber
\eeq
and consequently
\beq
\zeta\left(-1,{M^2\over2e{\mathcal E}}\right) &=&
-\,\frac{1}{2\pi^2}\sum_{n=1}^\infty
n^{-2}\,\cos\left(n\,{\pi M^2\over e{\mathcal E}}\right)
\nonumber\\
\zeta^{\,\prime}\left(-1,{M^2\over2e{\mathcal E}}\right) &=&
\frac{1}{2\pi^2}\,
[\,1-{\bf C}-\ln2\pi\,]\sum_{n=1}^\infty
n^{-2}\cos\left(n\,{\pi M^2\over e{\mathcal E}}\right)
\nonumber\\
&-&\frac{1}{2\pi^2}
\sum_{n=1}^\infty
n^{-2}\,(\ln n)\cos\left(n\,{\pi M^2\over e{\mathcal E}}\right)
\nonumber\\
&-& \frac{1}{4\pi}\,
\sum_{n=1}^\infty
n^{-2}\,\sin\left(n\,{\pi M^2\over e{\mathcal E}}\right)
\nonumber
\eeq
where ${\bf C}$ is the Euler-Mascheroni constant.
It turns out that the effective lagrangian
density in the four dimensional Minkowski spacetime 
is achieved under the inverse Wick rotation
${\mathcal E}\rightarrow -\,iE$ which yields
\beq
{\cal L}^{\rm M}_{\rm eff}(E) &=& 
{\cal L}^{\rm E}_{\rm eff}(-iE) = 
-\,\frac12\,E^2 - \left(\frac{M}{2\pi}\right)^2
\nonumber\\
&-& {e^2E^2\over2\pi^2}\left\lbrace
\left(1+\ln\frac{iM^2}{2eE}\right)
\zeta\left(-1,{iM^2\over2eE}\right)
- \zeta^{\,\prime}\left(-1,{iM^2\over2eE}\right)
\right\rbrace
\nonumber
\eeq
It follows therefrom that  we eventually find
\beq
\Im{\rm m}\,{\cal L}^{\rm M}_{\rm eff}(E)&=&
{e^2E^2\over8\pi^3}
\sum_{n=1}^\infty {1\over n^2}\,
\exp\left\{-\,n\,{\pi M^2\over eE}\right\}
\eeq
in perfect agreement with eq.~(\ref{instanton})
together with
\beq
\Re{\rm e}\,{\cal L}^{\rm M}_{\rm eff}(E)&=&
-\,{e^2E^2\over4\pi^4}\left({\bf C}+\ln\frac{\pi M^2}{eE}\right)
\sum_{n=1}^\infty {1\over n^2}\,
\cosh\left\{n\,{\pi M^2\over eE}\right\}
\nonumber\\
&-& {e^2E^2\over4\pi^4}\sum_{n=1}^\infty {\ln n\over n^2}\,
\cosh\left\{n\,{\pi M^2\over eE}\right\}
\eeq
As a consequence, the total rate of pairs production and/or annihilation
in the whole space and during all the time will be given by the 
manifestly Lorentz invariant expression
\beq
w&\equiv& 1-\exp\left\{-2{\mathcal V\,T}\;
\Im{\rm m}\,{\cal L}^{\,\rm M}_{\rm eff}({\mathcal F})\right\}
\nonumber\\
&=& 1-\exp\left(-\,{e^2{\mathcal F}\over4\pi^3}
\sum_{n=1}^\infty {1\over n^2}\,
\exp\left\{-\,{n\pi M^2\over e\,{\mathcal F}^{1/2}}\right\}\right)
\nonumber\\
&\simeq& {e^2{\mathcal F}\over4\pi^3}
\,\exp\left\{-\,{\pi M^2\over e\,{\mathcal F}^{1/2}}\right\}
\nonumber\\
{\mathcal F}&\equiv& -\,\frac12\,F_{\mu\nu}\,F^{\mu\nu}={\bf E}^2-{\bf B}^2
\eeq
when ${\mathcal G}\equiv\frac12\,\epsilon^{\mu\nu\rho\sigma}
F_{\mu\nu}\,F_{\rho\sigma}={\bf E}\cdot{\bf B}=0\,.$
As a matter of fact, once a uniform magnetic field is switched on, orthogonal to the
electrostatic field, we have to solve
the corresponding Dirac equation.
Instead of solving explicitely the above equation, we
can take profit of being within the
context of a relativistic theory, 
so that it is expedient to consider a new inertial
frame $K^{\,\prime}$ moving along the negative $OY$ axis with the velocity 
$v_2=-v\,,\ v>0$ with respect to the previously considered inertial
frame $K\,.$
Then the crossed magnetostatic
field does actually appear in $K^{\,\prime}$, {i.e.},
\[
{\bf E}^{\,\prime}=(E^{\,\prime}=\gamma E,0,0,)
\qquad\quad
{\bf B}^{\,\prime}=(0,0,\gamma v E=B^{\,\prime}\,)
\]
so that $v=(B^{\,\prime}/E^{\,\prime})\,.$
Notice, {\it en passant}, that by means of a Lorentz
transformation and just owing to the Lorentz invariance, we
shall be restricted to the case in which 
${\bf E}^{\,\prime}\cdot {\bf B}^{\,\prime}=0$,
namely electric and magnetic orthogonal fields, as well asin the first two sections
${\bf E}^{\,\prime\,2}>{\bf B}^{\,\prime\,2}$, that means 
relatively weak magnetostatic field, {viz.} $0<v<1\,.$
The spinor field in the $K^{\,\prime}$ reference frame can be obtained
in turn from the transformation law
$\Psi^{\,\prime}(x^{\,\prime})=\Lambda(v)\,\Psi(x)\,.$
It follows therefrom, taking eq.s~(\ref{polarization1}), (\ref{polarization2})
and (\ref{inelectron_r}) 
suitably into account, that the spinor solutions which describe an 
incoming electron, in the presence of crossed constant electromagnetic fields,
can be written as
\beq
&& [\,u_{\,{\bf p}\,,\,r}^{\,(-)}\,(t,{\bf r})\,]^{\,\prime}=[\,2eE\lambda(2\pi)^3\,]^{-1/2}\,
\exp\left\{i\,{\bf p}\cdot{\bf r} -\,{\pi\lambda/8}\right\}
\nonumber\\
&& \times\ \left\lbrace\Upsilon^{\,\prime}_r\,D_{\,i\lambda/2}\,(z_-) + 
\widetilde\Upsilon^{\,\prime}_r\,(1-i)\,{\textstyle\frac12}\,\lambda\,\sqrt{eE}\,D_{\,i\lambda/2-1}\,(z_-)
\right\rbrace
\nonumber
\label{inelectron4B}
\eeq
with
\beq
\Upsilon^{\,\prime}_r=\Lambda(v)\,\Upsilon_r\qquad 
\widetilde\Upsilon^{\,\prime}_r=\Lambda(v)\,\widetilde\Upsilon_r\qquad\quad
(\,r=1,2\,)
\eeq
and corresponding rather analogous expressions for all
other solutions of the Dirac equation. From the Lorentz
invariance of the effective lagrangian we immediately obtain
\beq
&& \Im{\rm m}\,{\cal L}^{\rm M}_{\rm eff}
({\bf E}^{\,\prime},{\bf B}^{\,\prime})\ =\nonumber\\
&& =\ {e^2\over8\pi^3}\,({\bf E}^{\,\prime\,2}-{\bf B}^{\,\prime\,2})
\sum_{n=1}^\infty n^{-2}\,
\exp\left\{-\,{n\pi M^2\over
e\,|\,{\bf E}^{\,\prime}-{\bf B}^{\,\prime}\,|}\right\}
\eeq
in accordance with the result of refs.~\cite{IZ}.

If, instead, in addition to the electrostatic field there is also a uniform
magnetic field not orthogonal to the electric field, 
then in a suitable Lorentz coordinate system we can always take ${\bf E}$
and ${\bf B}$ to be directed along the $OX$ axis. In such a circumstance,
it turns out that the imaginary part of the effective lagrangian
is provided by the celebrated Schwinger's formula \cite{nikishov,schwinger}
\beq
\Im{\rm m}\,{\cal L}^{\rm M}_{\rm eff}(E,B)&=&
{e^2BE\over8\pi^2}
\sum_{n=1}^\infty \frac{1}{n}\,
\exp\left\{-\,n\,{\pi M^2\over eE}\right\}\,
\coth\left(n\pi\,{B\over E}\right)
\nonumber
\eeq
\end{document}